\documentclass[a4paper,11pt]{article}
\usepackage[utf8]{inputenc}
\usepackage[english]{babel}
\usepackage{braket}
\usepackage{xspace}
\usepackage{bbm}
\usepackage{mathdots}
\usepackage{stackrel}
\usepackage[dvipsnames]{xcolor}
\usepackage{appendix}
\usepackage{hyperref}
\hypersetup{
 colorlinks=true,
 linkcolor=blue,
 anchorcolor = blue,
 citecolor = blue,
 filecolor = blue,
 urlcolor = blue
}
\usepackage{mathtools}
\usepackage{bbm}
\usepackage{comment}
\usepackage{subfigure}
\usepackage[T1]{fontenc}               
\usepackage[left=3cm,
            right=2.5cm,
            top=2.5cm,
            bottom=3cm
            ]{geometry}                
\usepackage{graphicx}
\usepackage{mathrsfs}
\usepackage{appendix}
\usepackage{fancyhdr}
\usepackage{amsmath,                   
            amssymb,                   
            amsthm}                    

\usepackage{setspace}
\usepackage{cite}
\usepackage{cancel}
\usepackage[margin=30pt, bf, font=small, center, justification=justified]{caption}[2004/07/16]

\usepackage{setspace} 

\usepackage{tikz}

\title{\bf Non-Abelian entanglement asymmetry in random states }

\author{Angelo Russotto$^{1}$, Filiberto Ares$^{1}$, and Pasquale Calabrese$^{1,2}$}

\date{}

\begin{document} 

\maketitle
{\small
\vspace{-5mm}  \ \\
{$^{1}$}  SISSA and INFN Sezione di Trieste, via Bonomea 265, 34136 Trieste, Italy\\[-0.1cm]
\medskip
{$^{2}$}  International Centre for Theoretical Physics (ICTP), Strada Costiera 11, 34151 Trieste, Italy\\[-0.1cm]
\medskip
}

\begin{abstract}
The entanglement asymmetry measures the extent to which a symmetry is broken within a subsystem of an extended quantum system.
Here, we analyse this quantity in Haar random states for arbitrary compact, semi-simple Lie groups, building on and generalising recent results obtained for the $U(1)$ symmetric case.
We find that, for any group, the average entanglement asymmetry vanishes in the thermodynamic limit when the subsystem is smaller than its complement.
When the subsystem and its complement are of equal size, the entanglement asymmetry jumps to a finite value, indicating a sudden transition of the subsystem from a fully symmetric state to one devoid of any symmetry. 
For larger subsystem sizes, the entanglement asymmetry displays a logarithmic scaling with a coefficient fixed by the dimension of the group. 
We also investigate the fluctuations of the entanglement asymmetry, which tend to zero in the thermodynamic limit. We check our findings against exact numerical calculations for the $SU(2)$ and $SU(3)$ groups. We further discuss their implications for the thermalisation of isolated quantum systems and black hole evaporation. 
\end{abstract}

\tableofcontents

\begin{section}{Introduction}

Random matrix theory was first introduced into physics by Wigner and Dyson~\cite{w-55, d-62, d-62-1} to model the energy spectrum of complex many-body quantum systems, such as heavy nuclei or molecules. Their essential idea was to describe overly complicated systems, for which an exact solution is unfeasible, as a black box in which a very large number of degrees of freedom interact according to unknown laws. Then, from a minimum set of assumptions and relying on typicality arguments, some generic, universal properties of the spectrum are predicted by a specific ensemble of random Hamiltonians conveniently chosen. Random matrices have been successfully applied to many other areas of physics, backed by a robust mathematical framework and a well-established set of methods that enable analytic calculations~\cite{m-04, f-10}. 
For example, we can mention
the characterisation of quantum chaos and the thermalisation of isolated quantum systems~\cite{bgs-84, d-91, s-94, akpr-16, klp-18}, transport in disordered media~\cite{b-97}, or the encoding of quantum information~\cite{cn-16}.

In the same spirit as Wigner and Dyson, a natural approach to understanding the properties of generic many-body quantum states is to exploit random matrix theory. This question has recently garnered significant attention. The common wisdom is that random quantum states, constructed using random matrices drawn from a particular ensemble, describe the qualitative behaviour of the excited eigenstates of typical Hamiltonians~\cite{vr-17, nwfs-18, fnsw-18, lg-19, bhkrv-22}. They are also expected to capture the universal features of states resulting from sufficiently chaotic dynamics. In this regard, the works~\cite{p-93, p-93-1} by Page on the black hole information paradox represent a turning point. By assuming that black holes follow a highly chaotic unitary dynamics, Page proposed to effectively describe them and the radiation emitted during their evaporation as a system of qubits in a random state. The central result of Page's papers is that, for a quantum system with an infinite number of degrees of freedom and divided into two parts $A$ and $B$, the average entanglement entropy over pure states distributed according to the Haar measure is the maximal one, 
\begin{equation}\label{eq:page_curve}
\mathbb{E}[S_1(\rho_A)]\simeq \min(\log d_A, \log d_B),
\end{equation}
where $d_A$ and $d_B$ are the dimensions of the Hilbert spaces of $A$ and $B$ respectively and  
$S_1(\rho)=-{\rm Tr}[\rho\log\rho]$ is the von Neumann entropy of the $A$'reduced density matrix $\rho_A$.  
Eq.~\eqref{eq:page_curve} is usually known as Page curve and has been proven in different ways~\cite{fk-94, s-95, s-96}. It implies that, for very large systems, typical quantum states have maximal bipartite entanglement. 

In black hole theory, the results obtained by Page are today a fundamental guiding principle in solving the information paradox~\cite{h-75, h-76}. In particular, through the replica wormhole approach, several recent works~\cite{aemm-19, penington-20, ahmst-20, pssy-22} have found in certain specific quantum gravity models that the entanglement entropy of the radiation satisfies the Page curve and information is not lost during the evaporation. Eq.~\eqref{eq:page_curve} also plays an essential role in understanding the thermalisation of isolated quantum systems following a chaotic evolution with no conservation laws, 
since the ensemble of Haar random states describes their late-time behaviour~\cite{akpr-16}.

The work of Page also paved the way for investigating other entanglement properties of generic many-body quantum systems, usually a very challenging task, by applying random matrix theory. For Haar random states, there are many results on their higher entanglement moments~\cite{csz-06,g-07,vpo-16,w-17}, entanglement spectrum~\cite{z-06,mbl-08,dffp-13}, mixed-state purity~\cite{depasqualeetal}, negativity~\cite{slkv-21}, or relative entropy~\cite{kf-21}.   Other ensembles of random states have been also considered. For example, the typical behaviour of non-interacting systems is described by random Gaussian states~\cite{vhbr-17, vhbr-18, hvrb-19, lrv-20, lrv-21}. The analogue of the Page curve for the latter has been obtained in Refs.~\cite{lcb-18, zlc-20, bhk-21, bp-21}.  It has also been calculated for the complex Ginibre ensemble~\cite{ckf-23}, which characterises non-Hermitian/dissipative chaotic quantum systems.
A very interesting aspect of random states is the analysis of their symmetries. One possibility is to consider an ensemble of random states with a global internal symmetry and investigate its effects on the average entanglement entropy. This has been done in Refs.~\cite{vr-17, lcb-18, bhkrv-22, bhk-21, gg-18} and, more generally, in~\cite{ypzrh-23} for systems with particle number conservation, and in~\cite{phfr-23} for non-Abelian symmetries. Another possibility is examining the symmetry resolution of entanglement~\cite{lr-14,gs-18,xas-18,cdm-21}, that is how entanglement distributes among the symmetry sectors, as has been done in random states both for Abelian~\cite{bd-19,lntt-22,mcp-22,g-24} and non-Abelian symmetry groups~\cite{bdk-24}. 

Equally relevant is the analysis of the broken symmetries. A crucial point, which has been largely overlooked until very recently, is that, in many-body quantum systems, the breaking of a symmetry inherently depends on the particular subsystem that we consider, due to the presence of long-range correlations that do not respect the symmetry. Therefore, a natural candidate to diagnose symmetry breaking is a quantity based on entanglement measures. This is the idea behind the entanglement asymmetry, which measures the degree to which a symmetry is broken in a part of the system in terms of the entanglement entropy.
This proposal emerged from different contexts~\cite{vawj-08,gms-09,chmp-20,chmp-21,amc-23}.
Using the definitions of \cite{amc-23}, the breaking of an arbitrary $U(1)$ symmetry in a system of qubits in Haar random states, identical to the one considered by Page, has been recently analysed in Ref.~\cite{ampc-24}. The main result of that work is the following. If we take an infinite system and we divide it into two parts $A$ and $B$, then the state of $A$ is $U(1)$ symmetric for $d_A<d_B$. At $d_A=d_B$, the subsystem $A$ undergoes a sharp transition to a state that breaks the symmetry when $d_A>d_B$. Surprisingly, the entanglement asymmetry displays a discontinuity at the transition point. As discussed in Ref.~\cite{ampc-24}, this result might have important consequences for the symmetries of the radiation emitted in the evaporation of a black hole, as well as for the symmetries of the local stationary state to which a generic isolated non-equilibrium system with no conserved charges relaxes. 
The goal of the present manuscript is to extend the results of Ref.~\cite{ampc-24} for the $U(1)$ entanglement asymmetry in Haar random states to any compact, semi-simple Lie group.

\textbf{Definitions and setup.} In this work, we consider a system of $L$ qudits that we split into two sets $A$
and $B$ of $\ell_A$ and $\ell_B=L-\ell_A$ qudits respectively. Therefore, the total Hilbert space
is $\mathcal{H}=\mathbb{C}^d\otimes\stackrel{L}{\cdots} \otimes \mathbb{C}^d$, of dimension $d^L$, 
and factorises into the Hilbert subspaces of $A$ and $B$, $\mathcal{H}=\mathcal{H}_A\otimes\mathcal{H}_B$,
of dimensions $d_A=d^{\ell_A}$ and $d_B=d^{\ell_B}$. The state of the total system 
$\ket{\Psi}\in\mathcal{H}$ is a Haar random state, obtained as $\ket{\Psi}=U\ket{0}$, where
$U$ is a $d^L\times d^L$ unitary matrix drawn from the Haar unitary ensemble and $\ket{0}$ is 
just a reference state. The state of one of the subsystems, for example $A$, is given 
by the reduced density matrix $\rho_A={\rm Tr}_B[U\ket{0}\bra{0}U^\dagger]$. The ensemble of
mixed states obtained by partially tracing a pure Haar random state is also known as Wishart-Laguerre
random matrix ensemble~\cite{f-10}. Haar random states are typically not symmetric under any group $G$. The goal of 
this paper is to measure how much a global internal symmetry corresponding to a compact Lie group $G$ is 
broken in the subsystem $A$ described by the state $\rho_A$ using the entanglement asymmetry. 

To define the entanglement asymmetry, we require that the unitary representation $U_g$ acting on $\mathcal{H}$ of any element $g$
of the group $G$ decomposes as $U_g=U_{g,A}\otimes U_{g,B}$, where $U_{g, A}$ and $U_{g, B}$ are 
unitary representations of $g$ in the subsystems $A$ and $B$. In that case, under the action of $g$,
the reduced density matrix $\rho_A$ transforms as $\rho_A\mapsto U_{g, A} \rho_A U_{g, A}^\dagger$.  
Therefore, $\rho_A$ is symmetric when it is invariant under this transformation, that is when $[\rho_A, U_{g, A}]=0$,
for any group element $g$. Based on this observation, from any reduced density matrix $\rho_A$ we can 
construct a symmetrised density matrix
\begin{equation}
\label{sym_rho}
    \rho_{A,G} = \frac{1}{\text{vol}\,G} \int_G \text{d}g \, U_{A,g} \rho_A U^{\dagger}_{A,g},
\end{equation}
where $\text{d}g$ is the invariant measure of the group $G$ and $\text{vol} \, G$ its volume. By construction, $\rho_{A, G}$ is symmetric, i.e. $[\rho_{A, G}, U_{g, A}]=0$, for any $g\in G$. Then the $n$ R\'enyi entanglement asymmetry is defined as~\cite{amc-23}
\begin{equation}
\label{defAsym}
    \Delta S_A^{(n)} = S_n(\rho_{A,G}) - S_n(\rho_A),
\end{equation}
that is the difference between the R\'enyi entropies, $S_n(\rho) = \frac{1}{1-n} \log \text{Tr}[\rho^n]$, of $\rho_{A, G}$ and $\rho_{A}$. A particular very relevant case is the limit $n\to 1$, in which the R\'enyi entropy gives the von Neumann entropy $S_1(\rho)=-{\rm Tr}[\rho\log\rho]$ and Eq.~\eqref{defAsym} becomes
\begin{equation}
\Delta S_A=S_1(\rho_{A, G})-S_1(\rho_A). 
\end{equation}
The main properties satisfied by the entanglement asymmetry, which make it a quantifier of symmetry breaking in $A$, are $\Delta S_A^{(n)}\geq 0$ and $\Delta S_A^{(n)}=0$ if and only if $\rho_A$ is symmetric, that is when $\rho_A=\rho_{A, G}$. The entanglement asymmetry has been recently analysed in various different settings: algebraic quantum field theory~\cite{chmp-20, chmp-21, magan2-21,magan-21,ch-22, bcklm-24}, with applications to gauge theories, non-invertible and higher form symmetries, conformal field theories~\cite{chch-23, cm-23,fadc-24,kmop-24,frc-24, bgs-24}, spin chains~\cite{lmac-24}, and matrix product states~\cite{cv-23}. The entanglement asymmetry is of particular interest in non-equilibrium setups, such as quantum quenches, to monitor in a subsystem the time evolution of a symmetry broken by the initial configuration but respected by the post-quench dynamics. Unexpectedly, it has been found that, under certain circumstances, the more the symmetry is initially broken, the faster is locally restored after the quench. This phenomenon, analogous to the classical Mpemba effect, has been observed in free and interacting integrable models~\cite{amc-23, makc-24, rkacmb-24,yca-24,yac-24,bkccr-24, klobas-24, fcb-24, rvc-24, cma-24, avm-24, carc-24} as well as in chaotic systems, such as random circuits~\cite{tcd-24, lzyz-24}, in the presence of disorder~\cite{lzyzy-24, hbz-24}, and in holographic CFTs~\cite{bgs-24}. The entanglement asymmetry has been also employed to explore the absence of dynamical symmetry restoration~\cite{amvc-23}, the effects of confinement~\cite{khoretal-24} and other out-of-equilibrium features of discrete~\cite{kmop-24, fac-23, mff-24} and non-internal symmetries, for example spatial translations~\cite{krb-24}. The R\'enyi entanglement asymmetry has also been measured and the quantum Mpemba effect was observed experimentally in an ion-trap quantum simulator~\cite{joshietal-24} employing the randomised measurement toolbox~\cite{elben-23}, in which random matrices also play a crucial role. With the different perspective of quantum resource theory, the same quantity has also been studied in Refs.~\cite{vawj-08,gms-09,t-19,ms-14, tfhtp-24}. 

Plugging the definition of $\rho_{A, G}$ in Eq.~\eqref{defAsym} and applying the right invariance of the Haar measure, it can be easily proven that, for integer R\'enyi index $n$, the entanglement asymmetry can be written as~\cite{cv-23, fadc-24}
\begin{equation}
\label{def_asym_chargedmom}
\Delta S^{(n)}_A = \frac{1}{1-n} \log \Bigg[\frac{1}{(\text{vol}\,G)^{n-1}} \int_{G^{n-1}} \text{d} \mathbf{g}\,  \frac{Z_n(\mathbf{g})}{Z_n(\mathbf{e})}\Bigg],
\end{equation}
where $G^{n-1} = G \times\stackrel{n-1}{\cdots}  \times G $, $\mathbf{g} = (g_1,\cdots,g_{n-1}) \in G^{n-1}$ and $\mathbf{e} = (e_1,\cdots,e_{n-1})$ with $e$ the identity element of the group $G$. Finally, $Z_n(\mathbf{g})$ are the charged moments of $\rho_A$
\begin{equation}
\label{charged_mom}
Z_n(\mathbf{g}) = \text{Tr}[\rho_A U_{A,g_1} \rho_A U_{A,g_1} \cdots \rho_A U_{A,g_{n-1}^{-1} \cdots g_1^{-1}}].
\end{equation}
Since the full system of qudits is in a Haar random state, we are thus interested in computing the average entanglement asymmetry $\mathbb{E}[\Delta S_A^{(n)}]$ over the states $\rho_A$ obtained by partially tracing a Haar random state. The average value that we will obtain will also be the typical entanglement asymmetry, in the sense that the fluctuations around the average value go to zero as the dimension of the total Hilbert space approaches infinity.

\textbf{Outline}. The paper is organised as follows. In Sec.~\ref{sec_abelian}, we review the results obtained in Ref.~\cite{ampc-24} for the average entanglement asymmetry of the Abelian $U(1)$ group. In doing so, we introduce a convenient graphical notation to determine the leading asymptotic behaviour of the asymmetry for large systems. Moreover, we discuss the fluctuations of the asymmetry around its average value. In Sec.~\ref{sec_su2}, we study the entanglement asymmetry for the non-Abelian group $SU(2)$. We find both the finite-size average $n=2$ Rényi asymmetry and its large system size behaviour for any $n$. In Sec.~\ref{sec_genG}, we generalise the later result to a generic semi-simple compact Lie group. We particularise it to the $SU(N)$ group and check against numerical calculations for the $SU(3)$ case.
Finally, the conclusions and some outlooks are presented in Sec.~\ref{sec_concl}.
\end{section}

\begin{section}{Review of the Abelian case}
\label{sec_abelian}
In this section, we review the known results on the entanglement asymmetry in Haar random states for the Abelian group $U(1)$ obtained in Ref.~\cite{ampc-24}. 
In this simpler case, we can consider a system made of qubits. Therefore, the local Hilbert space dimension is $d=2$ and the dimensions of the Hilbert spaces of the subsystems $A$ and $B$ are $d_A = 2^{\ell_A}$ and $d_B = 2^{\ell_B}$, respectively.  The local basis for the $k$-th qubit is given by two orthonormal states that we will denote as $\ket{0}_k$ and $\ket{1}_k$, $k=1,\dots, L$, following the notation of Ref.~\cite{ampc-24}. As already described in general terms, we prepare the total system in the pure state $\rho = U\ket{0}\bra{0}U^{\dagger}$ with $\ket{0} = \bigotimes_k \ket{0}_k$ and $U$ a $2^L\times 2^L$ random unitary matrix drawn from the Haar ensemble.
A global internal $U(1)$ symmetry is generated by a single charge that we choose to be $Q = \sum_{k=1}^L \ket{1}_k \bra{1}_k$, which counts the number of qubits in the state $\ket{1}_k$. Observe that $Q$ can be decomposed as $Q=Q_A+Q_B$ where $Q_A$ and $Q_B$ are the restriction of $Q$ to the subsystems $A$ and $B$. Therefore, the unitary representation of the elements of the $U(1)$ group in the subsystem $A$ is given by $U_{A, \alpha}=e^{i\alpha Q_A}$, where $\alpha\in[-\pi, \pi)$.
 
The first step to compute the corresponding average R\'enyi entanglement asymmetry $\mathbb{E}[\Delta S^{(n)}_A] $ consists in taking the average inside the logarithm of the Rényi entropy, i.e. 
\begin{equation}
\mathbb{E}[\log \text{Tr}[\rho^n_A]] \simeq \log \mathbb{E}[\text{Tr}[\rho^n_A]],
\end{equation}
for both $\rho_A$ and $\rho_{A,G}$.
As already discussed in~\cite{ampc-24}, this approximation has been verified numerically and the error committed is exponentially small in the system size $L$. We will later motivate this approximation further when discussing the behaviour of the fluctuations of the entanglement asymmetry. The advantage of this approximation is that we only need to compute the average value of the moments of $\rho_A$ and $\rho_{A, G}$, a much easier problem. In particular, specialising Eq.~\eqref{def_asym_chargedmom} for the case of the $U(1)$ group considered here, the average moments of $\rho_{A, G}$ can be calculated from
\begin{equation}
\label{avg_sym_rho}
    \mathbb{E}[\text{Tr}[\rho_{A,G}^n]] = \int_{-\pi}^{\pi}\frac{{\rm d}\alpha_1\cdots {\rm d}\alpha_{n-1}}{(2 \pi)^{n-1}} \mathbb{E}[Z_n(\boldsymbol{\alpha})],
\end{equation}
where $Z_n(\boldsymbol{\alpha})$ are the $U(1)$ charged moments
\begin{equation}
\label{abelian_chargedmom}
    Z_n(\boldsymbol{\alpha}) = \text{Tr}[\rho_A e^{i \alpha_1 Q_A} \cdots \rho_A e^{-i(\sum_{i=1}^{n-1}\alpha_i) Q_A}].
\end{equation}
The average charged moments $\mathbb{E}[Z_n(\boldsymbol{\alpha})]$ can now be computed by using standard methods discussed in the literature in the context of random quantum circuits, see e.g.~\cite{fknv-23, pv-22}. 

A useful representation of the charged moments~\eqref{abelian_chargedmom} is that of a ``multi-layer'' circuit obtained by stacking multiple replicas of the total density matrix $\rho$. To obtain it, we have to apply the Choi-Jamiolkowski mapping that allows us to write the density matrix $\rho$ as the vector $ | \rho \rangle\!\rangle \in \mathcal{H} \otimes \mathcal{H}$. In particular, for our Haar random state $\rho  = U\ket{0} \bra{0} U^{\dagger}$, we have $|\rho \rangle\!\rangle =(U \otimes U^*) \ket{0}^{\otimes 2}$. In this way, $\rho$ is obtained by stacking together $U$ and $U^*$. In the case of the charged moments~\eqref{abelian_chargedmom}, we just need to stack together $n$ copies of the density matrix, as we graphically represent in Fig.~\ref{fig:Z2abelian} for $n=2$, and join them by contracting $|\rho\rangle\!\rangle^{\otimes n}$ with some boundary state $|\hspace{-1.5pt}-\hspace{-1.2pt}+;\boldsymbol{\alpha}\rangle\!\rangle$ that implements the traces in $A$ and $B$ as well as the insertions $e^{i\alpha_j Q_A}$ between each copy,
\begin{equation}
\label{Zn_vectorized}
    Z_n(\boldsymbol{\alpha}) = \langle\!\langle-+;\boldsymbol{\alpha}|(U\otimes U^*)^{\otimes n} \ket{0}^{\otimes2 n}.
\end{equation}
\begin{figure}[t]
\includegraphics[scale = 0.2]{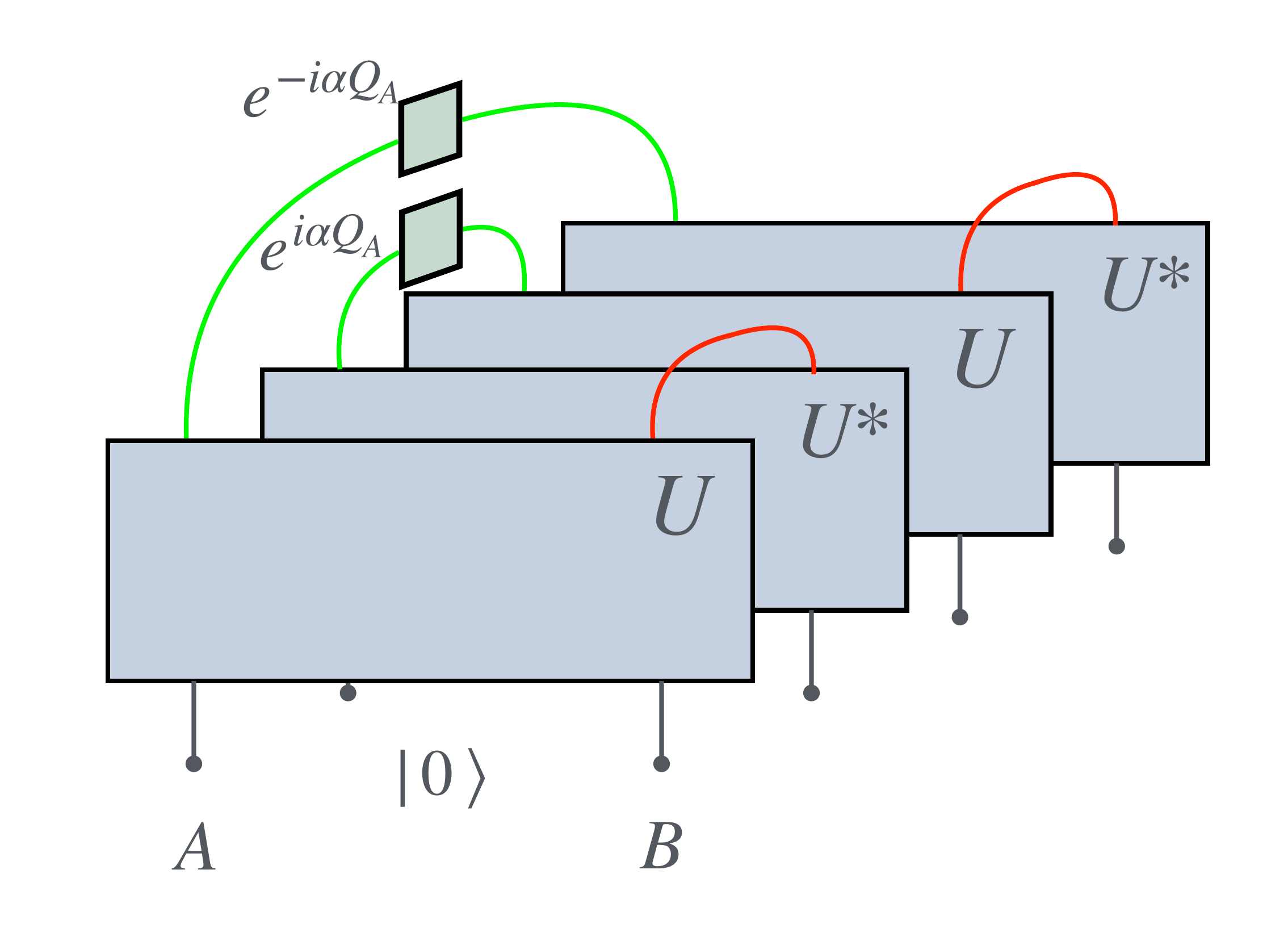}
\centering
\caption{Pictorial representation of the $n=2$ charged moment $Z_2(\boldsymbol{\alpha})$ in Eq.~\eqref{Zn_vectorized}. The left input leg represents the collection of indices belonging to the subsystem $A$, while the right leg corresponds to those of the complement $B$. The red and green contractions are implemented, respectively, by the boundary states $\bigotimes_k | +\rangle\!\rangle_k$ and $\bigotimes_k | -,\boldsymbol{\alpha}\rangle\!\rangle_k$ defined in Eqs.~\eqref{eq:plus_bs_ab} and \eqref{eq:minus_bs_ab}.}
\label{fig:Z2abelian}
\end{figure}
The boundary state is, in this case, given by
\begin{equation}
\label{boundarystate_u1}
    |\hspace{-1.5pt}-\hspace{-1.2pt}+;\boldsymbol{\alpha}\rangle\!\rangle = \bigotimes_{k \in A}|-;\boldsymbol{\alpha}\rangle\!\rangle_k  \bigotimes_{k \in B} |+\rangle\!\rangle_k.
\end{equation}
 For the qubits $k\in B$, the boundary state has to implement the partial trace of each replica of the density matrix $\rho$,  connecting the copies of $U$ and $U^*$ as indicated by the red contractions in Fig.~\ref{fig:Z2abelian}.
Therefore, one has 
\begin{equation}\label{eq:plus_bs_ab}
    |+\rangle\!\rangle_k = \sum_{\{a_j = 0, 1\}} \bigotimes_{j=1}^n (\ket{a_j}_k \otimes \ket{a_j}_k).
\end{equation}
Instead, for the qubits $k\in A$, the boundary state has to connect the copies of $U$ and $U^*$ as indicated by the green contractions in Fig. \ref{fig:Z2abelian},  implementing also the insertion of the operators $e^{i\alpha_j Q_A}$ between the replicas. Assuming $a_{n+1} \equiv a_1$ and $\alpha_n \equiv -\sum_{j=1}^{n-1}\alpha_j$, then
\begin{equation}\label{eq:minus_bs_ab}
    |-;\boldsymbol{\alpha} \rangle\!\rangle_k = \sum_{\{a_j = 0, 1\}} \bigotimes_{j=1}^n (\ket{a_j}_k \otimes e^{-i\alpha_j a_{j+1}} \ket{a_{j+1}}_k) ,
\end{equation}
where we used the fact that the charge operator $Q_A$ is diagonal in the basis $\{\ket{a_j}\}$. Such simplification will not occur when generalising the computation to a generic Lie group.

We are now in the position to compute the average of $Z_n(\boldsymbol{\alpha})$ over the Haar unitary ensemble as 
\begin{equation}
\label{EZn_u1}
    \mathbb{E}[Z_n(\boldsymbol\alpha)] = \langle\!\langle-+;\boldsymbol{\alpha}|\,\mathbb{E}[(U\otimes U^*)^{\otimes n}]\ket{0}^{\otimes2 n}.
\end{equation}
The average over the Haar ensemble of a tensor product of unitaries is a well-known result in random matrix theory~\cite{w-78,cs-06}
\begin{equation}\label{Haar_avg}
\mathbb{E}[U^{\otimes n} \otimes (U^*)^{\otimes n}] = \sum_{\sigma,\tau \in S_n} \text{Wg}(\sigma \tau^{-1}) |\sigma\rangle\!\rangle \langle\!\langle\tau|,
\end{equation}
where $S_n$ is the symmetric group and the state $|\sigma \rangle\!\rangle = \otimes_k |\sigma\rangle\!\rangle_k$ is 
\begin{equation}
    |\sigma \rangle\!\rangle_k = \sum_{\{a_j = 0\}}^{1} \bigotimes_{j=1}^n (\ket{a_j}_k \otimes \ket{a_{\sigma(j)}}_k).
\end{equation}
The Weingarten coefficients $\text{Wg}(\sigma)$ are rational functions of the elements of the symmetric group $S_n$. In this paper, we are interested in the asymptotic behaviour of the entanglement asymmetry for large system size $L$. In that limit, the Weingarten coefficients are of the form~\cite{cs-06}
\begin{equation}
\label{}
    \text{Wg}(\sigma) = \frac{1}{2^{n L}}\Bigg[\frac{\text{Moeb}(\sigma)}{2^{L |\sigma|}}+ \mathcal{O}\Bigg(\frac{1}{2^{L|\sigma|+2 L}}\Bigg)\Bigg],
\end{equation}
where $|\sigma|$ is the minimum number of transpositions to obtain the permutation $\sigma$. In particular, for the identity $\text{Id}\in S_n$, we have $|\text{Id}| = 0$, while for the rest of permutations $|\sigma| > 0$. Therefore, the coefficients of the terms $\sigma\neq \tau$ in Eq.~\eqref{Haar_avg} are exponentially smaller in $L$ than the ones of the terms $\sigma=\tau$. For large system sizes $L$, we can then neglect the terms $\sigma\neq \tau$ in Eq.~\eqref{Haar_avg}. The Möbius function $\text{Moeb}(\sigma)$ depends only on the cycle structure of $\sigma$. In our case, we only need to know that $\text{Moeb}(\text{Id}) = 1$. 
We can thus approximate the Weingarten coefficients in \eqref{Haar_avg} as $\text{Wg}(\sigma) = 2^{-n L}\delta_{\sigma,\text{Id}}$, up to exponentially small corrections in $L$~\cite{jyvl-20}. Applying~\eqref{Haar_avg} and  this approximation for $\text{Wg}(\sigma)$ in Eq.~\eqref{EZn_u1}, the average charged moments can be rewritten as 
\begin{equation}
    \mathbb{E}[Z_n(\boldsymbol{\alpha})] \simeq \frac{1}{2^{n L}} \sum_{\sigma \in S_n} \langle\!\langle -+;\boldsymbol{\alpha}| \sigma \rangle\!\rangle,
\end{equation}
where we took into account that $\langle\!\langle\sigma | 0\rangle^{\otimes 2n} = 1$ for any $\sigma$. Plugging in this last equation the explicit expression~\eqref{boundarystate_u1} of the boundary state $|\!-+;\boldsymbol{\alpha}\rangle\!\rangle$, we obtain
\begin{equation}\label{eq:av_ch_mom_ab_bs}
    \mathbb{E}[Z_n(\boldsymbol{\alpha})] \simeq \frac{1}{2^{n L}} \sum_{\sigma \in S_n} (\langle\!\langle-;\boldsymbol{\alpha}| \sigma\rangle\!\rangle_k)^{\ell_A} (\langle\!\langle+| \sigma\rangle\!\rangle_k)^{L-\ell_A}.
\end{equation}
Computing the overlaps appearing in the latter formula one gets~\cite{ampc-24}
\begin{equation}
    \mathbb{E}[Z_n(\boldsymbol{\alpha})] \simeq \frac{1}{2^{n L}} \sum_{\sigma \in S_n} \Big(\sum_{\{a_j\}} \prod_{j=1}^{n} \delta_{a_j,a_{\sigma(j)}} \Big)^{L-\ell_A}\Big(\sum_{\{a_j\}} \prod_{j=1}^{n} e^{i \alpha_j a_{j+1}}\delta_{a_{\sigma(j)},a_{j+1}}\Big)^{\ell_A}.
\end{equation}
To understand better this result and extract the leading order term at large $L$, it is useful to introduce a graphical notation which straightforwardly allows one to determine the contribution coming from each permutation $\sigma$. A similar diagrammatic approach has been discussed and used to study the entanglement negativity spectrum in random mixed states in \cite{slkv-21}. Here, we represent graphically the boundary states introduced in \eqref{boundarystate_u1} as 
\begin{center}
\begin{equation}
\label{diagram+state}
\begin{tikzpicture}[scale = 0.5,baseline=(current  bounding  box.center)]
\node at (-4,0){$|+\rangle\!\rangle_k = $};
\node at (-2,1) {\footnotesize$1$};
\node at (0,1.05) {\footnotesize$1'$};
\draw (0,0.5) arc[start angle=0, end angle=-180, radius=1];
\node at (1,1) {\footnotesize$2$};
\node at (3,1.05) {\footnotesize$2'$};
\draw (3,0.5) arc[start angle=0, end angle=-180, radius=1];
\draw (6,0.5) arc[start angle=0, end angle=-180, radius=1];
\node at (8,0) {$\cdots$};
\node at (10,1) {\footnotesize$n$};
\node at (12,1.1) {\footnotesize$n'$};
\draw (12,0.5) arc[start angle=0, end angle=-180, radius=1];
\node at (12.5,0) {,};
\end{tikzpicture}
\end{equation}
\end{center}

\begin{center}
\begin{equation}
\label{diagram-state}
\begin{tikzpicture}[scale = 0.5,baseline=(current  bounding  box.center)]
\node at (-4,0){$|-;\boldsymbol{\alpha}\rangle\!\rangle_k = $};
\draw[rounded corners=10pt] (-1.5,0.5) -- (-1.5,-1) -- (12,-1) -- (12,0.5);
\fill (5.25,-1) circle (4pt);
\node at (5.25,-1.5) {\footnotesize $\alpha_n$};
\draw (1.5,0.5) arc[start angle=0, end angle=-180, radius=1];
\fill (0.5,-0.5) circle (4pt);
\node at (0.5,0) {\footnotesize $\alpha_1$};
\draw (4.5,0.5) arc[start angle=0, end angle=-180, radius=1];
\fill (3.5,-0.5) circle (4pt);
\node at (3.5,0) {\footnotesize $\alpha_2$};
\node at (7,0) {$\cdots$};
\draw (11,0.5) arc[start angle=0, end angle=-180, radius=1];
\fill (10,-0.5) circle (4pt);
\node at (10,0) {\footnotesize $\alpha_{n-1}$};
\node at (12.5,0) {.};
\end{tikzpicture}
\end{equation}
\end{center}
Each line implements a contraction between the replicas of the qubit $k$. 
The presence of a dot in a contraction indicates the insertion of the operator $e^{i \alpha q_{A,k}}$ with $q_{A,k} = \ket{1}_k \bra{1}_k$. Analogously, the state $| \sigma \rangle\!\rangle_k$ can be thought of as a graph in which each line connects the replica $i$ and $\sigma(i)'$ of the qubit $k$. In this way, the contraction between $| \sigma \rangle\!\rangle_k$ and the boundary state amounts to simply counting the number of closed loops. The contribution of each loop is 
\begin{center}
\begin{equation}\label{eq:undressed_loop}
\begin{tikzpicture}[scale = 0.5,baseline=(current  bounding  box.center)]
    \draw (0,0) circle[radius = 1];
    \node at (2,0) {$= 2$ ,};
\end{tikzpicture}
\end{equation}
\end{center}
\begin{center}
\begin{equation}
\label{abelian_loop}
\begin{tikzpicture}[scale = 0.5,baseline=(current  bounding  box.center)]
    \draw (0,0) circle[radius = 1];
    \fill (0,-1) circle (4pt);
    \fill (0,1) circle (4pt);
    \node at (7,0) {$= \text{Tr}[e^{i \alpha_i q_{A}}e^{i \alpha_j q_{A}}] = 1 + e^{i (\alpha_i+\alpha_j)}$ ,};
    \node at (0,-1.5) {\footnotesize $\alpha_i$};
    \node at (0,1.5) {\footnotesize $\alpha_j$};
\end{tikzpicture}    
\end{equation}
\end{center}
where for the dressed loop we are using the fact that $q_A$ generates an Abelian group to simplify the expression for the trace of the product of two unitary representations of the group to the trace of one unitary representation with a parameter which is the sum of the two parameters. 

Coming back to Eq.~\eqref{eq:av_ch_mom_ab_bs}, the leading contribution to $\mathbb{E}[Z_n(\boldsymbol{\alpha})]$ in the large $L$ limit depends on whether $\ell_A<L/2$ or $\ell_A>L/2$. In the first case, the leading order term is determined by the permutation that maximises $\langle\!\langle +|\sigma \rangle\!\rangle_k$. This corresponds to the identity permutation, which gives the maximum possible number of loops, i.e. $n$ loops. Graphically
\begin{center}
\begin{equation}
\label{+sigmaid}
\begin{tikzpicture}[scale = 0.5,baseline=(current  bounding  box.center)]
\node at (-4,0){$\langle\!\langle+|\text{Id} \rangle\!\rangle_k = $};
\draw (-0.5,0) circle[start angle=0, end angle=-180, radius=1];
\draw (2.5,0) circle[start angle=0, end angle=-180, radius=1];
\node at (4.9,0) {$\cdots$};
\draw (7,0) circle[start angle=0, end angle=-180, radius=1];
\node at (8.5,0) {,};
\end{tikzpicture}
\end{equation}
\end{center}
\begin{center}
\begin{equation}
\label{-sigmaid}
\begin{tikzpicture}[scale = 0.5,baseline=(current  bounding  box.center)]
\node at (-4,0){$\langle\!\langle-;\boldsymbol{\alpha}|\text{Id}\rangle\!\rangle_k = $};
\begin{scope}[shift = {(0,0.5)}]
\begin{scope}[shift = {(0,-0.5)}]
\draw[rounded corners=10pt] (-1.5,0.5) -- (-1.5,-1) -- (12,-1) -- (12,0.5);
\fill (5.25,-1) circle (4pt);
\node at (5.25,-1.5) {\footnotesize $\alpha_n$};
\draw (1.5,0.5) arc[start angle=0, end angle=-180, radius=1];
\fill (0.5,-0.5) circle (4pt);
\node at (0.5,0) {\footnotesize $\alpha_1$};
\draw (4.5,0.5) arc[start angle=0, end angle=-180, radius=1];
\fill (3.5,-0.5) circle (4pt);
\node at (3.5,0) {\footnotesize $\alpha_2$};
\node at (6.8,0) {$\cdots$};
\draw (11,0.5) arc[start angle=0, end angle=-180, radius=1];
\fill (10,-0.5) circle (4pt);
\node at (10,0) {\footnotesize $\alpha_{n-1}$};
\node at (12.5,0) {.};
\end{scope}
\draw (-0.5,0) arc[start angle=0, end angle=180, radius=0.5];
\draw (2.5,0) arc[start angle=0, end angle=180, radius=0.5];
\draw (5.5,0) arc[start angle=0, end angle=180, radius=0.5];
\draw (9,0) arc[start angle=0, end angle=180, radius=0.5];
\draw (12,0) arc[start angle=0, end angle=180, radius=0.5];
\end{scope}
\end{tikzpicture}
\end{equation}
\end{center}
Neglecting the rest of the permutations in Eq.~\eqref{eq:av_ch_mom_ab_bs} and applying Eqs.~\eqref{+sigmaid} and \eqref{-sigmaid}, we conclude that the average charged moments behave as
\begin{equation}\label{avgzn+}
    \mathbb{E}[Z_n(\boldsymbol{\alpha})] \simeq 2^{(1-n) \ell_A}, \hspace{1cm} \frac{\ell_A}{L} < \frac{1}{2}.
\end{equation}
In the regime $\ell_A/L > 1/2$, we need instead to maximise $\langle\!\langle-;\boldsymbol{\alpha}| \sigma\rangle\!\rangle_k$ and, therefore, the number of dressed loops. This can be understood from the fact that the contribution of each loop is $2 \, e^{i \alpha/2} \cos (\alpha/2)$ according to Eq.~\eqref{abelian_loop}. The phase factor disappears when we multiply all the loops due to the neutrality condition $\sum_{j=1}^n \alpha_j = 0$. Thus, if a permutation $\sigma$ yields $M_{\sigma}$ dressed loops, its total contribution is $2^{M_{\sigma} \ell_A}f_{\sigma}(\boldsymbol{\alpha})^{\ell_A}$. What remains is the integration~\eqref{avg_sym_rho} over the group $U(1)^{n-1}$ of a function $f_{\sigma}(\boldsymbol{\alpha})^{\ell_A}$ such that $|f_{\sigma}(\boldsymbol{\alpha})|\leq 1$. This integral, performed using a saddle point approximation for large $\ell_A$, goes to zero algebraically in $\ell_A$ with a power $p_{\sigma}\geq 0$ which depends explicitly on the permutation $\sigma$. Summing up, each permutation gives a contribution $\sim 2^{M_\sigma \ell_A}\, \ell_A^{-p_{\sigma}}$. We can then conclude that the leading permutation is the one maximising $M_{\sigma}$. This is precisely the cyclic permutation $\nu(j) = j+1$; graphically
\begin{center}
\begin{equation}
\label{+sigmacycl}
\begin{tikzpicture}[scale = 0.5,baseline=(current  bounding  box.center)]
\node at (-4,0){$\langle\!\langle+|\nu\rangle\!\rangle_k = $};
\begin{scope}[shift = {(1,0)}]

\draw (-0.5,0) arc[start angle=0, end angle=-180, radius=1];
\draw (0.5,0) arc[start angle=0, end angle=180, radius=0.5];
\draw (2.5,0) arc[start angle=0, end angle=-180, radius=1];
\draw (3.5,0) arc[start angle=0, end angle=180, radius=0.5];
\node at (4.8,0) {$\cdots$};
\begin{scope}[shift = {(2,0)}]
\draw (5,0) arc[start angle=0, end angle=180, radius=0.5];
\draw (7,0) arc[start angle=0, end angle=-180, radius=1];
\node at (7.5,0) {,};
\end{scope}

\draw[rounded corners=10pt] (-2.5,0) -- (-2.5,1.5) -- (9,1.5) -- (9,0);
\end{scope}
\end{tikzpicture}
\end{equation}
\end{center}
\begin{center}
\begin{equation}
\label{-sigmacycl}
\begin{tikzpicture}[scale = 0.5,baseline=(current  bounding  box.center)]
\node at (-4,0){$\langle\!\langle-;\boldsymbol{\alpha}|\nu\rangle\!\rangle_k = $};
\begin{scope}[shift = {(0,0.)}]
\begin{scope}[shift = {(0,-0.5)}]
\draw[rounded corners=10pt] (-1.5,0.5) -- (-1.5,-1) -- (12,-1) -- (12,0.5);
\fill (5.25,-1) circle (4pt);
\node at (5.25,-1.5) {\footnotesize $\alpha_n$};
\draw (1.5,0.5) arc[start angle=0, end angle=-180, radius=1];
\fill (0.5,-0.5) circle (4pt);
\node at (0.5,0) {\footnotesize $\alpha_1$};
\draw (4.5,0.5) arc[start angle=0, end angle=-180, radius=1];
\fill (3.5,-0.5) circle (4pt);
\node at (3.5,0) {\footnotesize $\alpha_2$};
\node at (6.8,0.4) {$\cdots$};
\draw (11,0.5) arc[start angle=0, end angle=-180, radius=1];
\fill (10,-0.5) circle (4pt);
\node at (10,0) {\footnotesize $\alpha_{n-1}$};
\node at (12.5,0.5) {.};
\end{scope}
\draw[rounded corners=10pt] (-1.5,0) -- (-1.5,1.5) -- (12,1.5) -- (12,0);
\draw (1.5,0) arc[start angle=0, end angle=180, radius=1];
\draw (4.5,0) arc[start angle=0, end angle=180, radius=1];
\draw (11,0) arc[start angle=0, end angle=180, radius=1];
\end{scope}
\end{tikzpicture}
\end{equation}
\end{center}
By neglecting the other terms in Eq.~\eqref{eq:av_ch_mom_ab_bs} and calculating the contribution of the loops above with Eqs.~\eqref{eq:undressed_loop} and \eqref{abelian_loop}, one finds the average charged moments behave as
\begin{equation}
\label{avgzn-}
    \mathbb{E}[Z_n(\boldsymbol{\alpha})] \simeq 2^{(1-n)(L-\ell_A)} \prod_{j = 1}^{n} \cos\bigg(\frac{\alpha_j}{2}\bigg)^{\ell_A}, \hspace{1cm} \frac{\ell_A}{L} > \frac{1}{2}.
\end{equation}

 From Eqs.~\eqref{avgzn+} and \eqref{avgzn-}, we can compute the asymptotic entanglement asymmetry for large $L$ inserting them in~\eqref{avg_sym_rho} and solving the integral by a saddle point approximation. The final result is
\begin{equation}\label{eq:asymp_beh_asymm_ab}
    \mathbb{E}[\Delta S_A^{(n)}] \simeq \begin{cases}
        0, & \ell_A < L/2, \\
        \frac{1}{2} \log(\ell_A \pi n^{1/(n-1)}/2) & \ell_A > L/2. 
    \end{cases}
\end{equation}
In the limit $n\to 1$, this result leads to the von Neumann entanglement asymmetry 
\begin{equation}
\mathbb{E}[\Delta S_A] \simeq 1/2 \log(\ell_A \pi /2) + 1/2,\quad \ell_A>L/2.
\end{equation}

Eq.~\eqref{eq:asymp_beh_asymm_ab} is the main result of Ref.~\cite{ampc-24}. When $\ell_A < L/2$, the asymmetry vanishes in the limit $L\to\infty$, indicating that the reduced density matrix $\rho_A$ is on average $U(1)$ symmetric in that regime. As discussed in~\cite{ampc-24}, this property can be understood through the Hayden-Preskill decoupling inequality~\cite{hp-07},  which states that 
\begin{equation}
\label{dec_ineq}
 \mathbb{E}\big[||\rho_A - \frac{\mathbb{I}}{2^{\ell_A}}||_1\big]^2 \leq 2^{L-2 \ell_A},
\end{equation}
where $||\,.\,||_1$ is the $L_1$ norm. The decoupling inequality tells us that, if $\ell_A < L/2$, the reduced density matrix is on average exponentially close to the normalised identity matrix (the infinite temperature state) in the $L\to \infty$ limit. In fact, according to Page's result \eqref{eq:page_curve}, the average entanglement entropy is maximal for $\ell_A<L/2$ in the thermodynamic limit. Since the identity matrix respects any symmetry, one expects from the inequality~\eqref{dec_ineq} that $\rho_A$ is on average symmetric with respect to any group when $\ell_A<L/2$ and $L\to\infty$, as we have just proved in Eq.~\eqref{eq:asymp_beh_asymm_ab} for an arbitrary $U(1)$ group. For this reason, we can anticipate that the asymmetry of a subsystem $\ell_A<L/2$ and large $L$ remains zero when we consider any other group. But the most surprising feature of the result in Eq.~\eqref{eq:asymp_beh_asymm_ab} is the presence of a sharp transition in the entanglement asymmetry at $\ell_A = L/2$ in which it jumps from zero to a finite value. That is, the subsystem $A$ transitions from a symmetric state to a non-symmetric one when $\ell_A>L/2$. This transition cannot be predicted with the decoupling inequality~\eqref{dec_ineq}. In the following sections, we investigate this abrupt jump for different non-Abelian groups.\\

\textbf{Fluctuations.} We conclude the analysis of the Abelian case with a discussion of the fluctuations of the entanglement asymmetry. We consider the variance of the von Neumann asymmetry,
\begin{equation}\label{eq:var_vn_asymm}
\text{Var}[\Delta S_A]=\mathbb{E}[\Delta S_A^2]-\mathbb{E}[\Delta S_A]^2.
\end{equation}
To obtain it, we need to compute the average of $\Delta S_A^2$, which is given by
\begin{equation}
\begin{split}
    \mathbb{E}[\Delta S^2_A] &= \mathbb{E}[(S_1(\rho_{A,G})-S_1(\rho_A))^2] = \\
    &= \mathbb{E}[S_1(\rho_{A,G})^2] + \mathbb{E}[S_1(\rho_A)^2]-2\,\mathbb{E}[S_1(\rho_{A,G})S_1(\rho_A)].
\end{split}
\end{equation}
Let us focus on the first term since all of the other are just particular cases in which we substitute a set of dressed contractions with normal contractions when we replace $\rho_{A,G}$ by $\rho_A$. The expectation value of $S(\rho_{A,G})^2$ can be obtained as 
\begin{equation}
    \mathbb{E}[S_1(\rho_{A,G})^2] = \lim_{n,m\to 1} \partial_n \partial_m \mathbb{E}[\text{Tr}[\rho_{A,G}^n]\text{Tr}[\rho_{A,G}^m]].
\end{equation}
As we did in Eq.~\eqref{avg_sym_rho}, each trace can be expressed as the integral of the corresponding charged moments
\begin{equation}
    \mathbb{E}[\text{Tr}[\rho_{A,G}^n]\text{Tr}[\rho_{A,G}^m]] = \int  \frac{{\rm d}\boldsymbol{\alpha} {\rm d}\boldsymbol{\beta}}{(2\pi)^{n+m-2}}  \mathbb{E}[Z_n(\boldsymbol{\alpha}) Z_m(\boldsymbol{\beta})].
\end{equation}
Using the same formalism discussed above, cf. Eq.~\eqref{Zn_vectorized}, the average of the product of two charged moments is given by
\begin{equation}\label{eq:av_prod_ch_mom}
\mathbb{E}[Z_n(\boldsymbol{\alpha})Z_m(\boldsymbol{\beta})]=\langle\!\langle B;\boldsymbol{\alpha}, \boldsymbol{\beta}|\mathbb{E}[(U\otimes U^*)^{\otimes (n+m)}] \ket{0}^{\otimes2(n+m)}
\end{equation}
where the boundary state $|B;\boldsymbol{\alpha},\boldsymbol{\beta} \rangle\!\rangle = \otimes_k |B;\boldsymbol{\alpha},\boldsymbol{\beta} \rangle\!\rangle_k $ is graphically
\begin{center}
\begin{equation}
|B;\boldsymbol{\alpha},\boldsymbol{\beta} \rangle\!\rangle_k = \begin{cases}
  \begin{tikzpicture}[scale = 0.38,baseline=(current  bounding  box.center)]
\node at (19,0) {$k\in B$,};
\node at (19,-3) {$k\in A$.};
\begin{scope}[shift = {(0.55,0)},scale = 1.35]    
      \node at (-2,1) {\footnotesize$1$};
      \node at (0,1.05) {\footnotesize$1'$};
\draw (0,0.5) arc[start angle=0, end angle=-180, radius=1];
\node at (1,1) {\footnotesize$2$};
\node at (3,1.05) {\footnotesize$2'$};
\draw (3,0.5) arc[start angle=0, end angle=-180, radius=1];
\node at (4,1) {\footnotesize$3$};
\node at (6,1.05) {\footnotesize$3'$};
\draw (6,0.5) arc[start angle=0, end angle=-180, radius=1];
\node at (8,0) {$\cdots$};
\node at (10,1) {\footnotesize$n+m$};
\node at (12.1,1.075) {\footnotesize$n'+m'$};
\draw (12,0.5) arc[start angle=0, end angle=-180, radius=1];
 \end{scope} 

\begin{scope}[scale = 1.08]
\begin{scope}[shift = {(-0.5,-3)},scale = 1]
    \draw[rounded corners=10pt] (-1.5,0.5) -- (-1.5,-1) -- (7,-1) -- (7,0.5);
\fill (2.75,-1) circle (4pt);
\node at (2.75,-1.5) {\footnotesize $\alpha_n$};
\draw (1.5,0.5) arc[start angle=0, end angle=-180, radius=1];
\fill (0.5,-0.5) circle (4pt);
\node at (0.5,0.4) {\footnotesize $\alpha_1$};
\node at (2.75,0) {$\cdots$};
\draw (6,0.5) arc[start angle=0, end angle=-180, radius=1];
\fill (5,-0.5) circle (4pt);
\node at (5,0.4) {\footnotesize $\alpha_{n-1}$};
\end{scope}
\begin{scope}[shift = {(+8.5,-3)}]
    \draw[rounded corners=10pt] (-1.5,0.5) -- (-1.5,-1) -- (7,-1) -- (7,0.5);
\fill (2.75,-1) circle (4pt);
\node at (2.75,-1.5) {\footnotesize $\beta_n$};
\draw (1.5,0.5) arc[start angle=0, end angle=-180, radius=1];
\fill (0.5,-0.5) circle (4pt);
\node at (0.5,0.4) {\footnotesize $\beta_1$};
\node at (2.75,0) {$\cdots$};
\draw (6,0.5) arc[start angle=0, end angle=-180, radius=1];
\fill (5,-0.5) circle (4pt);
\node at (5,0.4) {\footnotesize $\beta_{n-1}$};
\end{scope}
\end{scope}
\end{tikzpicture} 
    \\
\end{cases}
\end{equation}
\end{center}

Applying in Eq.~\eqref{eq:av_prod_ch_mom} the same approximation for the Weingarten formula~\eqref{Haar_avg} in the large $L$ limit as before, $\text{Wg}(\sigma) \simeq 2^{-(n+m) L}\delta_{\sigma,\text{Id}}$, we find
\begin{equation}
\label{expZZ}
    \mathbb{E}[Z_n(\boldsymbol{\alpha}) Z_m(\boldsymbol{\beta})] \simeq \frac{1}{2^{(n+m)L}} \sum_{\sigma \in S_n}  \prod_{k=1}^{L} \langle\!\langle B;\boldsymbol{\alpha},\boldsymbol{\beta}| \sigma \rangle\!\rangle_k.
\end{equation}
The identification of the leading contribution to \eqref{expZZ} follows the same approach as in the computation of $\mathbb{E}[Z_n(\boldsymbol{\alpha})]$. We have to look for the permutation that maximises the number of loops in the sites belonging to $B$ or $A$ whether $\ell_A<L/2$ or $\ell_A>L/2$ respectively.
For $\ell_A<L/2$, the leading permutation is again the identity. While for $\ell_A > L/2$ the permutation that yields the largest number of dressed loops is the product of the cyclic permutation $\nu(j) = j+1$ among the first $n$ legs and the same cyclic permutation among the remaining $m$ legs.
In both regimes, these permutations do not connect the first $n$ legs with the remaining $m$ ones. This implies that the leading contribution to the expectation value of the product of charged moments factorises as $\mathbb{E}[Z_n(\boldsymbol{\alpha})Z_m(\boldsymbol{\beta})] \simeq \mathbb{E}[Z_n(\boldsymbol{\alpha})]\mathbb{E}[Z_m(\boldsymbol{\beta})]$ for large $L$. At the level of the moments of the symmetrised density matrix $\rho_{A, G}$ we have, consequently, that
\begin{equation}
\label{factorization_mom}
     \mathbb{E}[\text{Tr}[\rho_{A,G}^n]\text{Tr}[\rho_{A,G}^m]] \simeq \mathbb{E}[\text{Tr}[\rho_{A,G}^n]] \, \mathbb{E}[\text{Tr}[\rho_{A,G}^m]].
\end{equation}
This means that, in the limit of large system size $L$, $\mathbb{E}[\Delta S_A^2] = (\mathbb{E}[\Delta S_A])^2$. Therefore, the variance~\eqref{eq:var_vn_asymm} of the von Neumann entanglement asymmetry goes to zero. This result allows us to conclude that the average value obtained from Eq.~\eqref{eq:asymp_beh_asymm_ab} is also the typical value of the entanglement asymmetry.

Furthermore, the factorisation~\eqref{factorization_mom} justifies the first approximation done at the beginning of this section, i.e. $\mathbb{E}[\log \text{Tr}[\rho^n_{A,G}]] \simeq \log \mathbb{E}[\text{Tr}[\rho^n_{A,G}]]$, for calculating the average R\'enyi entanglement asymmetry. In fact, since
\begin{equation}
\label{replica_trick}
    \mathbb{E}[\log \text{Tr}[\rho^n_{A,G}]] = \lim_{k\to 0} \partial_k \mathbb{E}[(\text{Tr}[\rho^n_{A,G}])^k],
\end{equation}
in the limit of large $L$, applying the factorisation~\eqref{factorization_mom}, we have $\mathbb{E}[(\text{Tr}[\rho^n_{A,G}])^k] \simeq (\mathbb{E}[\text{Tr}[\rho^n_{A,G}]])^k$, up to exponentially small corrections in $L$. Plugging the latter result in~\eqref{replica_trick}, we directly prove the self-averaging property of the moments of the symmetrised density matrix, $\mathbb{E}[\log \text{Tr}[\rho^n_{A,G}]] \simeq \log \mathbb{E}[\text{Tr}[\rho^n_{A,G}]]$, in agreement with the numerical observations of Ref.~\cite{ampc-24}. Using the factorisation~\eqref{factorization_mom}, we can also show that the variance $\text{Var}[\Delta S_A^{(n)}]$ of the R\'enyi entanglement asymmetry tends to zero with the total system size $L$.
A similar argument, but applied to the study of typical relative entropy in random states, has already been considered in~\cite{kf-21}.

As a numerical consistency check, we can compute with exact diagonalisation the variance of the entanglement asymmetry for small system sizes $L$. In the left panel of Fig.~\ref{fig:vNfluctuations}, we show the average value of $\sqrt{\text{Var}[\Delta S_A]}$ over a finite number of samples $N$ up to $L = 10$. We observe that, even for these small sizes, the variance exhibits a monotonic decrease towards zero as $L$ increases. Using exact diagonalisation, we can also visualise the probability distribution function of the entanglement asymmetry. As shown in the right panel of Fig.~\ref{fig:vNfluctuations}, when we fix $L$ and vary $\ell_A$ from $1$ to $L$, the probability distribution changes from having a significant skewness to a more symmetric shape. In particular, the sharp drop for $\ell_A = 1$ at $\Delta S_A<\mathbb{E}[\Delta S_A]$ 
is due to the fact that the entanglement asymmetry is a positive-definite quantity. Moreover, the probability distribution is not only 
more symmetric as $\ell_A$ increases but it is also wider, since the fluctuations of the asymmetry are bigger for larger $\ell_A$ and fixed $L$ as we can also see in the left panel.
Unfortunately, a full analytical understanding of the properties of the probability distribution of the entanglement asymmetry is far from trivial, and we consider it beyond the scope of this work.

\begin{figure}[t]
\begin{subfigure}
    \centering
    \raisebox{-9.9mm}{\includegraphics[width=.5\linewidth]{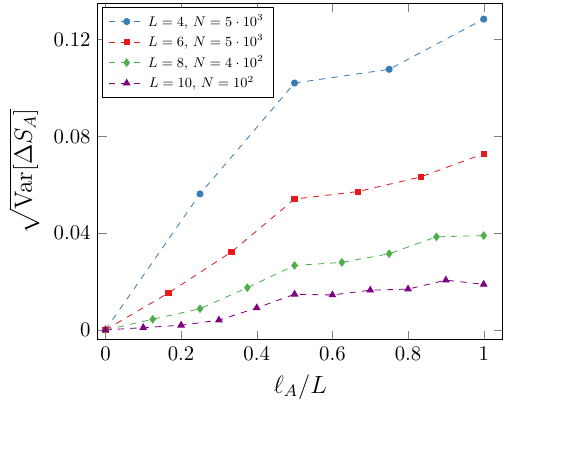}}
\end{subfigure}
\begin{subfigure}
    \centering
    \includegraphics[width=.42\linewidth]{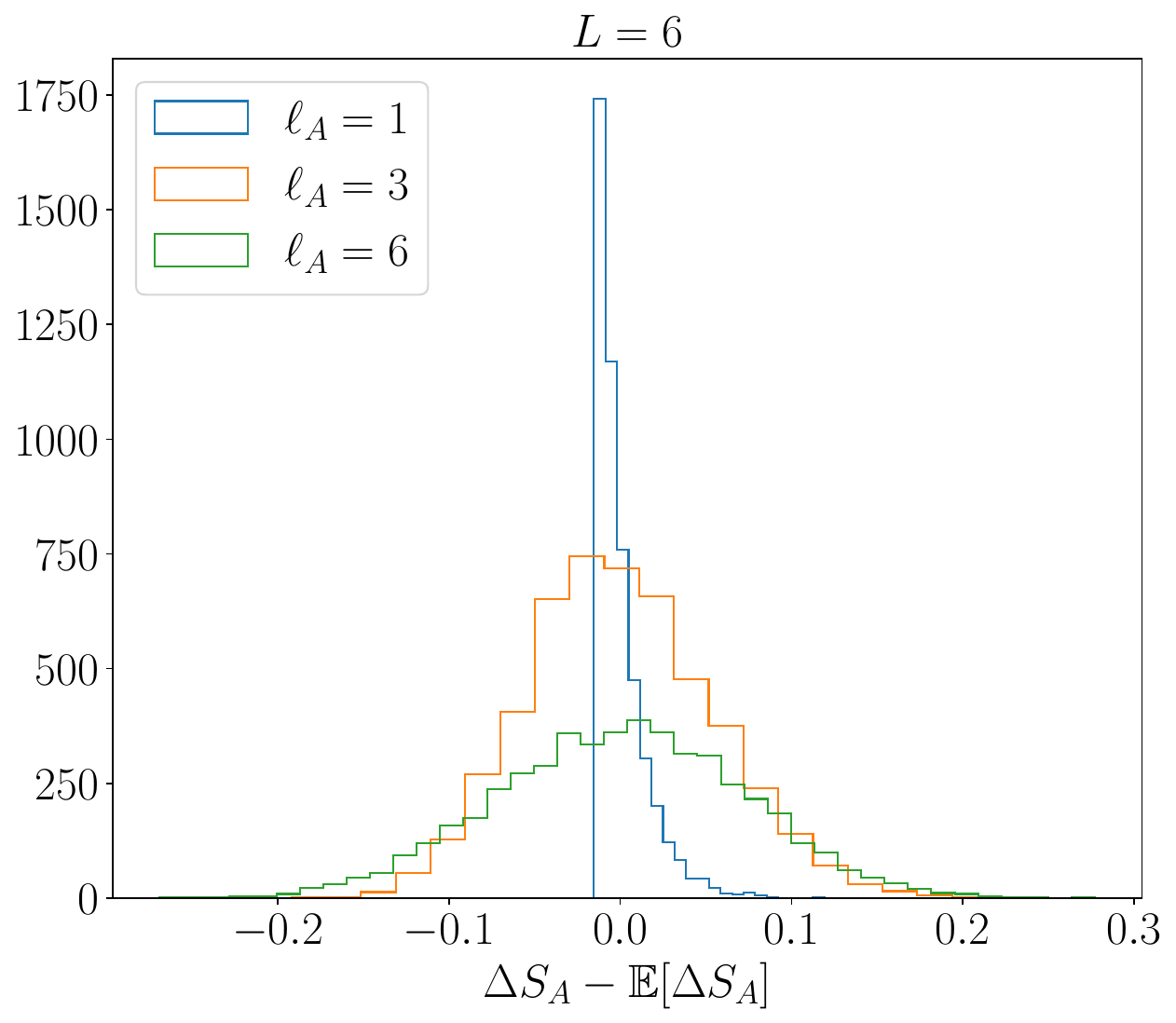}
\end{subfigure}
\caption{Left panel: Square root of the variance of the $U(1)$ von Neumann entanglement asymmetry as a function of the subsystem size $\ell_A$ for different sizes $L$ of the total system. 
The symbols are the variance computed numerically using exact diagonalisation over a finite number $N$ of samples of Haar random states. The dashed lines join the numerical data as guides for the eye. As $L$ increases the variance goes to zero independently of $\ell_A$ as predicted by the factorisation property of the expectation value~\eqref{factorization_mom} in the large $L$ limit. Right panel: Histogram of the occurrences of the $U(1)$ von Neumann entanglement asymmetry for total system of $L=6$ qubits. As the subsystem size $\ell_A$ varies from $1$ to $L$, the skewness of the distribution changes, becoming more symmetric and wider. The entanglement asymmetry has been calculated exactly using exact diagonalisation. The number of samples is $N = 5\cdot 10^3$.}
\label{fig:vNfluctuations}
\end{figure}

\end{section}
\begin{section}{Entanglement asymmetry for the $SU(2)$ group}
\label{sec_su2}

In this section, we extend the previous results for the Abelian Lie group $U(1)$ to the non-Abelian Lie group $SU(2)$. The analytical computation of the entanglement asymmetry for this specific group will provide the necessary insights for the generalisation to the case of a generic semi-simple compact Lie group, discussed in the next section.

As in the $U(1)$ case, we consider a system of $L$ qubits. As local basis for each qubit $k =1, \cdots, L$ we can use the eigenstates $\ket{\uparrow}_k$ and $\ket{\downarrow}_k$ of the Pauli matrix $\sigma_{z}^{(k)}$. We prepare the total system in the pure state $\ket{\uparrow} \equiv \bigotimes_{k=1}^L \ket{\uparrow}_k$ and then we apply a $2^L \times 2^L$ unitary matrix drawn from the Haar random ensemble. The reduced density matrix of a subsystem of $\ell_A$ qubits is therefore given by $\rho_A = \text{Tr}_B [U \ket{\uparrow} \bra{\uparrow} U^{\dagger}]$. We are interested in computing the average entanglement asymmetry $\mathbb{E}[\Delta S^{(n)}_A]$ associated with the $SU(2)$ group generated by the global charges $Q_{a} = \sum_{k= 1}^{L} \sigma_a^{(k)}/2\,$ for $a = x,y,z$, which form a basis of the corresponding $\mathfrak{su}(2)$ Lie algebra. Therefore, a unitary representation of this $SU(2)$ group on the total Hilbert space is given by $U_{g(\vec{x})}= e^{i \vec{x}\cdot \vec{Q}}$, where $\vec{x}=(x, y, z)\in \mathbb{R}^3$, with the constraint $|\vec{x}|\in[0, 2\pi]$, are a set of local coordinates of the $\mathfrak{su}(2)$ Lie algebra. The charges $Q_a$ are local in the sense that they decompose into the contribution from $A$ and the complement $B$ as $Q_a = Q_{A,a} + Q_{B,a}$ $\forall a$. This implies that $U_g$ satisfies the required factorisation property $U_g=U_{g, A}\otimes U_{g, B}$, with $U_{g(\vec{x}), A}= e^{i\vec{x}\cdot \vec{Q}_A}$ and $U_{g(\vec{x}), B}= e^{i\vec{x}\cdot \vec{Q}_B}$, to define the symmetrised reduced density matrix $\rho_{A,G}$ as in Eq.~\eqref{sym_rho}.

To proceed with the computation of $\mathbb{E}[\Delta S_A^{(n)}]$, we need to assume, as in the $U(1)$ case, the self-averaging property of the moments of the (symmetrised) reduced density matrix, i.e. $\mathbb{E}[\log\text{Tr}[\rho_{A,G}^n]] \simeq \log \mathbb{E}[\text{Tr}[\rho_{A,G}^n]]$. However, as it will become clear in the following, the discussion on the fluctuations of the entanglement asymmetry for the Abelian case can be straightforwardly applied in this case, as the diagrammatic procedure introduced in the previous section remains almost unchanged. We can thus safely bring the expectation value inside the logarithm. Applying then Eq.~\eqref{def_asym_chargedmom}, we can write the average Rényi entanglement asymmetry in terms of the previous local set of coordinates for $\mathfrak{su}(2)$ as
\begin{equation}
\label{avg_asym}
\mathbb{E}[\Delta S^{(n)}_A] \simeq \frac{1}{1-n} \log \Bigg[\frac{1}{(\text{vol}\,SU(2))^{n-1}} \int_{B^{2(n-1)}} \text{d} \mathbf{g}\,  \frac{\mathbb{E}[Z_n(\mathbf{g})]}{\mathbb{E}[Z_n(\mathbf{e})]}\Bigg],
\end{equation}
where $B^{2(n-1)}=B^2\times \stackrel{n-1}{\cdots}\times B^2$ and $B^2=\{\vec{x}\in \mathbb{R}^3\,|\, |\vec{x}|\leq 2\pi\}$. Here $\mathbf{g}=(g(\vec{x}_1),\dots, g(\vec{x}_{n-1}))$ and 
\begin{equation}\label{eq:ch_mom_su2}
\mathbb{E}[Z_n(\mathbf{g})] = \mathbb{E}[\text{Tr}[\rho_A U_{g(\vec{x}_1)} \cdots \rho_A U_{g(\vec{x}_n)}]],
\end{equation}
with $g(\vec{x}_n)=g(\vec{x}_{n-1})^{-1}\cdots g(\vec{x}_1)^{-1}$.
In this local set of coordinates and using the same normalisation convention of~\cite{cdm-21}, the $SU(2)$ Haar measure reads
\begin{equation}\label{eq:haarSU2_cart}
    \text{d}g(x,y,z) = \sqrt{2} \bigg(\frac{\sin( \sqrt{x^2+y^2+z^2}/2)}{\sqrt{x^2+y^2+z^2}}\bigg)^2\text{d}x\text{d}y\text{d}z.
\end{equation}
Changing to spherical coordinates $(x,y,z) = (\alpha \sin \beta \cos \gamma,\alpha \sin \beta \sin \gamma, \alpha \cos\beta)\equiv \alpha \,\hat{n}(\beta, \gamma)$, one has
\begin{equation}
\label{haarSU2}
    \text{d}g(\alpha,\beta,\gamma) = \sqrt{2}\sin^2 \frac{\alpha}{2} \sin \beta \text{d}\alpha \text{d}\beta \text{d} \gamma.
\end{equation}
In terms of them, the volume of the group can be directly obtained from
\begin{equation}
\label{volSU2}
    \text{vol}\, SU(2) = \int_{SU(2)} \text{d}g = \int_{|\Vec{x}|\leq 2 \pi} \text{d}g(\alpha,\beta,\gamma) = 2^{5/2} \pi^2.
\end{equation}

To compute the average charged moments, we can proceed as in the Abelian case and vectorise the total density matrix $\rho$ as $|\rho \rangle\!\rangle =(U \otimes U^*) \ket{\uparrow}^{\otimes 2}$. Then Eq.~\eqref{eq:ch_mom_su2} can be thought of as a replicated quantum circuit as the one in Fig.~\ref{fig:Z2abelian} contracted with a boundary state,
\begin{equation}
\label{EZ}
\mathbb{E}[Z_n(\mathbf{g})] = \langle\!\langle-+;\mathbf{g}| \mathbb{E}[U^{\otimes n} \otimes (U^*)^{\otimes n}] \ket{\uparrow}^{\otimes 2 n}.
\end{equation}
The contractions between the different replicas are the same as in the Abelian case, graphically represented in Fig.~\ref{fig:Z2abelian} for $n=2$, with the only difference that the operators inserted between the copies of $\rho_A$ are now $U_{g(\vec{x}_j), A}$. Therefore, the boundary state
$|-+;\mathbf{g}\rangle\!\rangle$ is of the same form,
\begin{equation}
|-+;\mathbf{g}\rangle\!\rangle = \bigotimes_{k \in A} |-;\mathbf{g}\rangle\!\rangle_k \bigotimes_{k \in 
B} |+\rangle\!\rangle_k,
\end{equation}
where the part for the qubits $k\in B$ remains untouched, cf. Eq.~\eqref{eq:plus_bs_ab},
\begin{equation}
\label{+state}
|+\rangle\!\rangle_k = \sum_{\{s_j=\uparrow,\downarrow\}} \bigotimes_{j=1}^n (\ket{s_j}_k \otimes \ket{s_j}_k ), 
\end{equation}
while the new operators $U_{g(\vec{x}_j), A}$ between the replicated qubits $k\in A$ are implemented as 
\begin{equation}
\label{-state}
|-;\mathbf{g}\rangle\!\rangle_k =  \sum_{\{s_j,s'_j=\uparrow,\downarrow\}} \bigotimes_{j=1}^n (\ket{s_j}_k \otimes (\mathrm{U}_{g_j})_{s_j',s_{j+1}} \ket{s_j'}_k),
\end{equation}
where $s_{n+1} \equiv s_1$. Here $\mathrm{U}_{g_j}$ is the two-dimensional unitary fundamental representation of $g(\vec{x}_j) \in SU(2)$ acting on the $k-$th qubit, i.e. $\mathrm{U}_{g_j}=e^{i\vec{x}_j\cdot \vec{Q}^{(k)}}$, with $Q_a^{(k)}=\sigma_a^{(k)}/2$, $a=x, y, z$. Observe that, unlike in the Abelian case~\eqref{eq:minus_bs_ab}, the elements of the local basis $\ket{\uparrow}_k$, $\ket{\downarrow}_k$ are not eigenstates of $\mathrm{U}_{g_j}$.

We can now apply in Eq.~\eqref{EZ} the formula~\eqref{Haar_avg}, which gives the average of the tensor product of Haar random unitary matrices. As in the $U(1)$ case, we can neglect the off-diagonal terms $\sigma\neq \tau$ approximating the Weingarten coefficients as $\text{Wg}(\sigma) = 2^{-n L}\delta_{\sigma,\text{Id}}$. In this way, we find the following expression for the average $SU(2)$ charged moments,
\begin{equation}
\label{resEZ}
\mathbb{E}[Z_n(\mathbf{g})] \simeq \frac{1}{2^{nL}} \sum_{\sigma \in S_n} \Big(\sum_{\{s_j=\uparrow,\downarrow\}} \prod_{j=1}^{n} \delta_{s_j,s_{\sigma(j)}} \Big)^{L-\ell_A}\Big(\sum_{\{s_j=\uparrow,\downarrow\}} \prod_{j=1}^{n} (\mathrm{U}_{g_j})_{s_{\sigma(j)},s_{j+1}}\Big)^{\ell_A},
\end{equation}
up to exponentially small corrections in $L$.\\

\textbf{Finite size analysis for $n = 2$}. Before studying the asymptotics for large system sizes of the entanglement asymmetry, it is worth investigating its finite-size behaviour. In fact, for $n=2$, the result in Eq.~\eqref{resEZ} simplifies considerably. In that case, there are only two possible permutations and the charged moment is 
\begin{equation}
\begin{split}
\mathbb{E}[Z_2(g)] &\simeq 4^{-\ell_A} \text{Tr}[\mathrm{U}^{\dagger}_g \mathrm{U}_g]^{\ell_A} + 2^{-L-\ell_A}\text{Tr}[\mathrm{U}_g]^{\ell_A}\text{Tr}[\mathrm{U}_g^{\dagger}]^{\ell_A} =\\
&\simeq 2^{-\ell_A} + 2^{-L+\ell_A}\left(\cos\frac{\alpha}{2}\right)^{2\ell_A}.
\end{split}
\end{equation}
In the second equality, we used the property that the unitary two-dimensional matrix $\mathrm{U}_g$ can be written as $\mathrm{U}_{g(\alpha,\beta,\gamma)} = \cos(\alpha/2) \mathbb{I} + i \hat{n}(\beta,\gamma)\cdot \Vec{\sigma} \sin(\alpha/2)$ in spherical coordinates. Plugging it into Eq.~\eqref{avg_asym}, 
\begin{equation}
\mathbb{E}[\Delta S^{(2)}_A] \simeq -\log\Big[\frac{1}{2^{-\ell_A} + 2^{-L+\ell_A}}\Big(2^{-\ell_A} + 2^{-L+\ell_A} \int_{0}^{2 \pi} \text{d}\alpha \frac{1}{\pi} \sin^2 \frac{\alpha}{2} \Big(\cos^{2}\frac{\alpha}{2}\Big)^{\ell_A} \Big)\Big],
\end{equation}
we obtain the average $n=2$ Rényi asymmetry
\begin{equation}
\label{DS2_analytic}
\mathbb{E}[\Delta S^{(2)}_A] \simeq -\log\left[ \frac{1}{1 + 2^{-L + 2\ell_A}} \left(1 + \frac{2^{-L}}{\ell_A + 1} \frac{(2 \ell_A)!}{\ell_A!^2}\right)\right].
\end{equation}

We can compare this result with the analogous one obtained in Ref.~\cite{ampc-24} for the $U(1)$ group,
\begin{equation}
    \mathbb{E}[\Delta S^{(2)}_A]|_{U(1)} \simeq -\log \left[\frac{1}{1 + 2^{-L + 2\ell_A}}\left( 1+2^{-L} \frac{(2 \ell_A)!}{\ell_A!^2}\right) \right].
\end{equation}
We observe that, independently of the value of $\ell_A/L$, we have $\mathbb{E}[\Delta S^{(2)}_A|_{SU(2)}]>\mathbb{E}[\Delta S^{(2)}_A|_{U(1)}]$. This is a very natural result as $SU(2)$ is a larger group containing $U(1)$ as a subgroup.

In Fig.~\ref{fig:DS2picture}, we check numerically Eq.~\eqref{DS2_analytic}. We compare our analytical prediction (solid lines) and the average R\'enyi entanglement asymmetry over a finite ensemble of Haar random states calculated using exact diagonalisation (symbols) for different system sizes $L$. We observe that, already for $L=4$, there is a good agreement between the numerical results obtained by computing numerically $\mathbb{E}[\log \text{Tr}\rho_{A,G}^n]$ and the analytical result obtained within the self-averaging approximation $\mathbb{E}[\log \text{Tr}\rho_{A,G}^n] \simeq \log \mathbb{E}[\text{Tr} \rho_{A,G}^n]$.\\ 

\begin{figure}[t]
\includegraphics[width=0.6\textwidth]{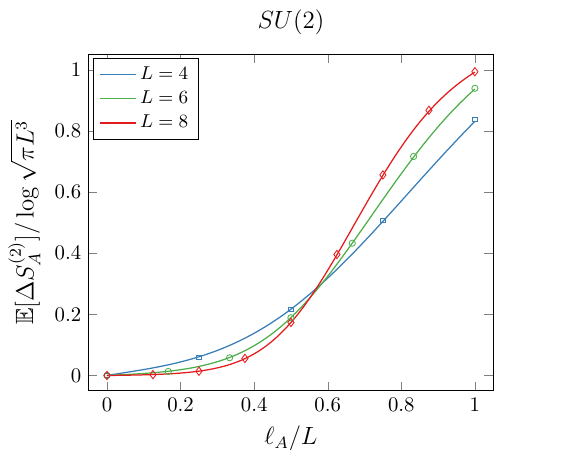}
\centering
\caption{Average $n=2$ $SU(2)$ Rényi asymmetry for Haar random states as a function of the subsystem size $\ell_A$ for different total system sizes $L$. The solid lines correspond to the analytical result in Eq.~\eqref{DS2_analytic}. The symbols are the numerically exact results obtained using exact diagonalisation by averaging $\Delta S^{(2)}_A$ over $ \sim 10^3$ random states in the Haar ensemble. This is enough to make the statistical uncertainty on the average smaller than the size of the symbol (for $L=8$, $10^2$ random states are enough to make the statistical error smaller than $\sim 2\cdot10^{-3}$). These results validate the self-averaging approximation employed to compute the average entanglement asymmetry also in the non-Abelian case.}
\label{fig:DS2picture} 
\end{figure}

\textbf{Large $L$ behaviour for generic $n$}.
The asymptotic behaviour of the charged moments~\eqref{resEZ} in the large $L$ limit can be studied using the diagrammatic approach introduced in the previous section. The boundary states \eqref{+state}-\eqref{-state} have the exact same structure as in the Abelian case, therefore they are described by the diagrams \eqref{diagram+state}-\eqref{diagram-state}. The only difference is that the dressed contractions contain now the non-diagonal and non-commuting operators $\mathrm{U}_{g_j}$. Hence, the simplification in~\eqref{abelian_loop} is not valid here. The rule for the contractions of dressed loops is now given by
\begin{center}
\begin{equation}
\label{non_abelian_loop}
\begin{tikzpicture}[scale = 0.5]
    \draw (0,0) circle[radius = 1];
    \fill (0,-1) circle (4pt);
    \fill (0.866,0.5) circle (4pt);
    \fill (-0.866,0.5) circle (4pt);
    \node at (4.5,0) {$= \text{Tr}[\mathrm{U}_i \mathrm{U}_j \mathrm{U}_k]$ ,};
    \node at (0,-1.5) {\footnotesize $i$};
    \node at (1.25574, 0.725) {\footnotesize $j$};
    \node at (-1.25574, 0.725) {\footnotesize $k$};
\end{tikzpicture}    
\end{equation}
\end{center}
where, to lighten the notation, we write $\mathrm{U}_j \equiv \mathrm{U}_{g_j}$. Notice, however, that this contraction is always of the form $\text{Tr}[\Tilde{\mathrm{U}}]$ with $\Tilde{\mathrm{U}}\in SU(2)$ the product of the unitary matrices entering in the loop, $\Tilde{\mathrm{U}}=\mathrm{U}_i \mathrm{U}_j \mathrm{U}_k \dots$. Therefore, it can always be written as $\text{Tr}[\Tilde{\mathrm{U}}]=2 \cos(\Tilde{\alpha}/2)$ where $\Tilde{\alpha}$ is an analytic function of the Lie algebra coordinates $\vec{x}_i, \vec{x}_j, \vec{x}_k,\dots $ associated to the operators $\mathrm{U}_i, \mathrm{U}_j, \mathrm{U}_k, \dots$. Therefore, as in the Abelian case, we expect that the contraction of a permutation $\sigma$ with the boundary state $|-;\mathbf{g} \rangle\!\rangle$ behaves, for large $\ell_A$, as $\sim 2^{M_{\sigma} \ell_A}\,\ell_A^{-p_{\sigma}}$, with $M_{\sigma}$ the number of dressed loops. The leading contribution for $\ell_A > L/2$ is thus given by the cyclic permutation $\nu(j)=j+1$ which maximises $M_{\sigma}$, and corresponds to the diagrams~\eqref{+sigmacycl}-\eqref{-sigmacycl}. On the other hand, for $\ell_A<L/2$, the leading contribution is given by the identity permutation $\text{Id}(j)=j$, which corresponds to the diagrams~\eqref{+sigmaid}-\eqref{-sigmaid}.
Hence, applying the rule~\eqref{non_abelian_loop} to these diagrams, we have for large $L$
\begin{equation}
\label{asymp_avgZn}
\mathbb{E}[Z_n(\mathbf{g})] \stackrel{L \to \infty}{\simeq} \begin{cases}
2^{(1-n)\ell_A}, & \ell_A < L/2 \\
2^{(1-n)L} 2^{-\ell_A} \big(\prod_{j=1}^{n-1}\text{Tr}[\mathrm{U}_j]^{\ell_A} \big) \text{Tr}[\mathrm{U}^{\dagger}_{n-1} \cdots \mathrm{U}^{\dagger}_{1}]^{\ell_A}, & \ell_A > L/2
\end{cases}.
\end{equation}

Plugging this result into Eq.~\eqref{avg_asym}, we obtain immediately $\mathbb{E}[\Delta S_A^{(n)}] \simeq 0 $ when $\ell_A < L/2$, as expected from the decoupling inequality~\eqref{dec_ineq}. Instead, for $\ell_A>L/2$, we need to analyse the large $\ell_A$ asymptotic behaviour of the following integral
\begin{equation}\label{eq:saddle_su2_eq}
\frac{\mathbb{E}[\text{Tr}[\rho_{A,G}^n]]}{\mathbb{E}[\text{Tr}[\rho_A^n]]} \simeq \frac{2^{-n\ell_A}}{(\text{vol}\, SU(2))^{n-1}}\int_{B^{2(n-1)}} \text{d}\mathbf{g}\,  \text{Tr}[\mathrm{U}^{\dagger}_{n-1} \cdots \mathrm{U}^{\dagger}_{1}]^{\ell_A} \prod_{j=1}^{n-1}\text{Tr}[\mathrm{U}_j]^{\ell_A},
\end{equation}
where we used $\mathbb{E}[Z_n(\mathbf{e})] = 2^{(1-n)(L-\ell_A)}$. This analysis can be done with a saddle-point approximation. The integrand possesses saddle points at the points $(\vec{x}_1,\dots,\vec{x}_{n-1})$ for which  $\mathrm{U}_j = \pm \mathbb{I}\,$, $j = 1,\cdots, n-1$, and, consequently, $\mathbb{E}[Z_n(\mathbf{g})] = \mathbb{E}[Z_n(\mathbf{e})]$. The behaviour of the integrand is the same around all of these saddle points and therefore we can approximate Eq.~\eqref{eq:saddle_su2_eq} as $2^{n-1}$ times the integral on a neighbourhood $I(\vec{0})$ of $(\vec{x}_1, \dots, \vec{x}_{n-1})=(\vec{0}, \dots, \vec{0})$, at which $\mathrm{U}_j = \mathbb{I}$ for all $j$,
\begin{equation}
\label{integralsaddle}
   \frac{\mathbb{E}[\text{Tr}[\rho_{A,G}^n]]}{\mathbb{E}[\text{Tr}[\rho_A^n]]} \simeq \frac{2^{n-1} 2^{-n \ell_A}}{(\text{vol}\,SU(2))^{n-1}} \int_{I(\vec{0})} \text{d}\mathbf{g} \exp \Big[\ell_A\big( \sum_{j=1}^{n-1} \log \text{Tr}[\mathrm{U}_j] + \log \text{Tr}[\mathrm{U}^{\dagger}_{n-1} \cdots \mathrm{U}^{\dagger}_{1}]\big)\Big].
\end{equation}
We now introduce spherical coordinates and use the decomposition
$\mathrm{U}_j = \cos(\alpha_j/2) \mathbb{I} + i \hat{n}(\beta_j, \gamma_j) \cdot \Vec{\sigma} \sin (\alpha_j/2)$ to compute the quadratic expansion of the argument of the exponential near $\alpha_j = 0$. In this way, we find
\begin{equation}
\log \text{Tr}[\mathrm{U}_j] \simeq \log 2 - \frac{\alpha_j^2}{8}.
\end{equation}
For the term $\text{Tr}[\mathrm{U}^{\dagger}_{n-1} \cdots \mathrm{U}^{\dagger}_{1}]$, if we also apply the property $\text{Tr}[\sigma_j\sigma_{j'}] = 2 \delta_{jj'}$, we eventually obtain
\begin{equation}
\log \text{Tr}[\mathrm{U}^{\dagger}_{n-1} \cdots \mathrm{U}^{\dagger}_{1}] \simeq \log 2 - \frac{1}{8} \sum_{j,j'}  \Vec{x}_j \cdot \Vec{x}_{j'},
\end{equation}
where $\Vec{x}_j = \alpha_j \hat{n}(\beta_j, \gamma_j)$. Plugging these expansions into Eq.~\eqref{integralsaddle} and using the explicit expression of the Haar measure~\eqref{eq:haarSU2_cart} in our local coordinates,
we have
\begin{equation}\label{eq:saddle_int_su2_1}
    \frac{\mathbb{E}[\text{Tr}[\rho_{A,G}^n]]}{\mathbb{E}[\text{Tr}[\rho_A^n]]} \simeq 2^{n-1} \int _{I(\vec{0})} \prod_{j=1}^{n-1}\Bigg[\frac{\sqrt{2}}{\text{vol}\,SU(2)}\Big(\frac{\sin |\Vec{x}_j|/2}{|\Vec{x}_j|}\Big)^2 {\rm d} \Vec{x}_j\Bigg]e^{-\frac{\ell_A}{8}\big(\sum_{j=1}^{n-1} \Vec{x}_j^2 + \sum_{j,j'} \Vec{x}_j \cdot \Vec{x}_{j'}\big)}.
\end{equation}
Finally, extending the domain of integration to $\mathbb{R}^{3(n-1)}$ and evaluating the Haar measure at $|\Vec{x}| = 0$, the $3(n-1)$-fold integral~\eqref{eq:saddle_int_su2_1} factorises into three identical $(n-1)$-fold Gaussian integrals
\begin{equation}\label{eq:gauss_int_su2}
    \frac{\mathbb{E}[\text{Tr}[\rho_{A,G}^n]]}{\mathbb{E}[\text{Tr}[\rho_A^n]]} \simeq \frac{1}{(2^3 \pi^2)^{n-1}} \Bigg[\int_{\mathbb{R}^{n-1}} \text{d}x_1\cdots {\rm d}x_{n-1}\,\, e^{-\frac{\ell_A}{8}\big(\sum_{j=1}^{n-1} x_j^2 + \sum_{j,j'} x_j x_{j'}\bigl)}\Bigg]^3 .
\end{equation}
The latter integral can be evaluated by standard means, yielding the following result for the average Rényi entanglement asymmetry:
\begin{equation}\label{eq:av_asymm_su2}
    \mathbb{E}[\Delta S^{(n)}_A] \simeq \begin{cases}
        0, & \ell_A < L/2, \\
        \frac{3}{2} \log \big(\ell_A \pi^{1/3}n^{\frac{1}{n-1}}/2\big), & \ell_A > L/2.
    \end{cases}
\end{equation}
The limit $n\to1$ gives the von Neumann entanglement asymmetry,
\begin{equation}
\mathbb{E}[\Delta S_A] \simeq \frac{3}{2} \log (\ell_A \pi^{1/3}/2) + \frac{3}{2},\quad \ell_A > L/2. 
\end{equation}
Compared to the Abelian case~\eqref{eq:asymp_beh_asymm_ab}, the coefficient of the logarithmic term in the regime $\ell_A>L/2$ has changed from $1/2$ to $3/2$, hinting at the possibility that the numerator of the logarithmic prefactor is related to the dimension of the group considered, as we prove in the next section.

The arguments employed in Sec.~\ref{sec_abelian} to show that the variance of the $U(1)$ entanglement asymmetry tends to zero in the limit $L\to \infty$ do not rely on the specific properties of the operators $U_{g, A}$ inserted between the copies of $\rho_A$ in the charged moments. We can therefore apply them directly to the $SU(2)$ case. This allows us to conclude that the result in Eq.~\eqref{eq:av_asymm_su2} is also the typical entanglement asymmetry for an arbitrary $SU(2)$ group. As a numerical check, in the left panel of Fig.~\ref{fig:asym2SU2fluct}, we have calculated the variance of $\Delta S^{(2)}_A$ using exact diagonalisation as a function of $\ell_A/L$ for increasing sizes $L$ of the system. We can see that the fluctuations of the asymmetry decrease as $L$ increases.
In the right panel of Fig.~\ref{fig:asym2SU2fluct}, we show the probability distribution of the $n=2$ R\'enyi entanglement asymmetry for $SU(2)$, as we did in the right panel of Fig.~\ref{fig:vNfluctuations} for the $U(1)$ group. To construct it, we took $N=2\cdot 10^3$ Haar random states and calculate their asymmetry with exact diagonalisation.

\begin{figure}[t]
\begin{subfigure}
    \centering
    \raisebox{-1.989mm}{\includegraphics[width=.5\linewidth]{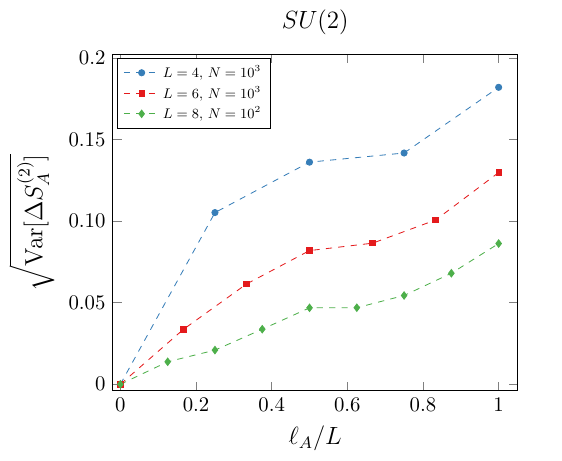}}
\end{subfigure}
\begin{subfigure}
    \centering
    \includegraphics[width=.405\linewidth]{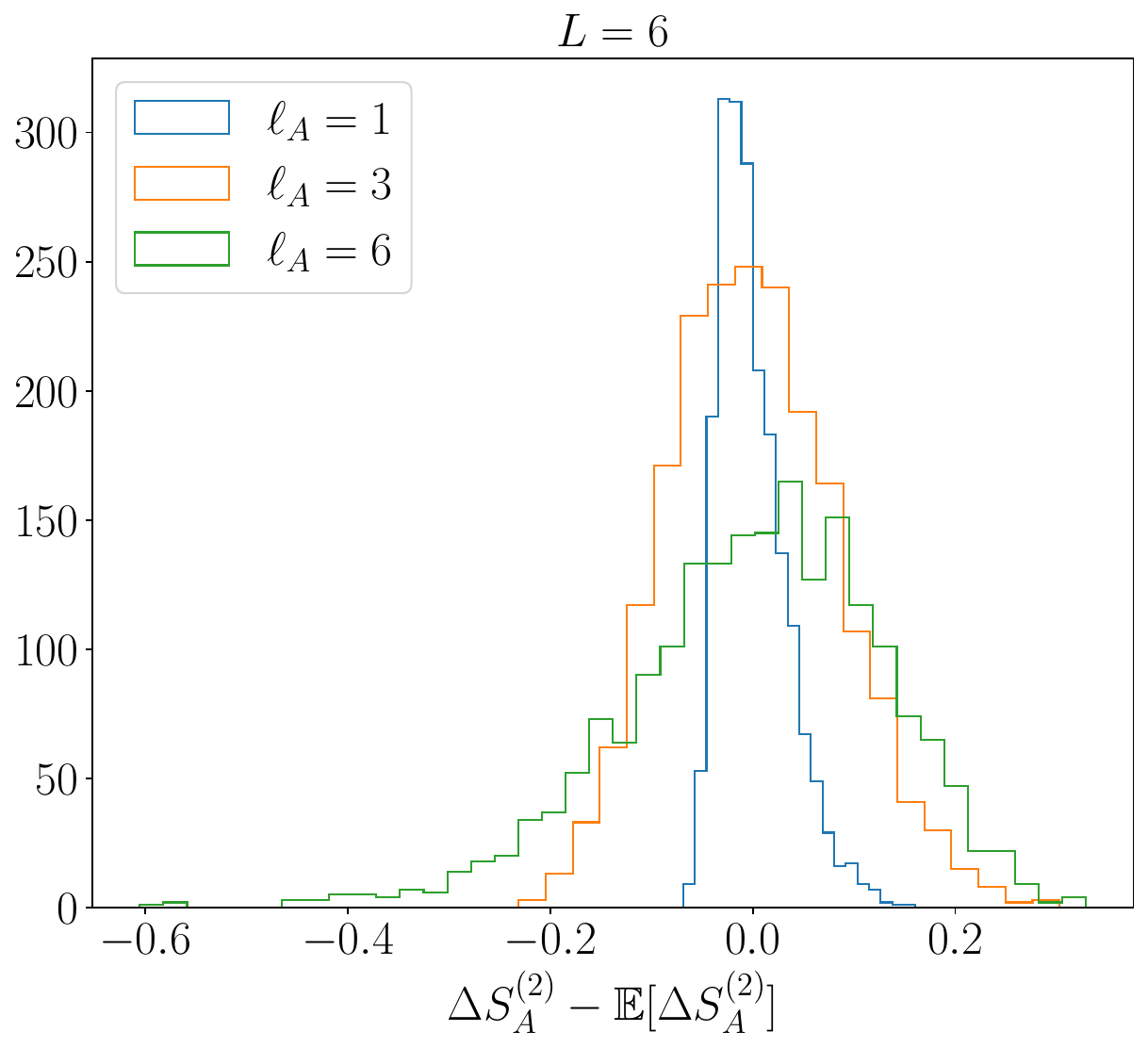}
\end{subfigure}
\caption{Same as in Fig.~\ref{fig:vNfluctuations} but for the $n = 2$ Rényi entanglement asymmetry of the $SU(2)$ group. We observe that, also in this case, the fluctuations around the average value decrease monotonically as we increase $L$. The number of samples for each histogram in the left panel is $N = 2\cdot10^3$.}
\label{fig:asym2SU2fluct}
\end{figure}

\end{section}
\begin{section}{Entanglement asymmetry for a generic compact Lie group $G$}
\label{sec_genG}

Finally, in this section, we calculate the main result of this paper which is the average entanglement asymmetry of a generic semi-simple compact Lie group $G$.

Let us consider a system of $L$ qudits with local Hilbert space of dimension $d$. We then take an irreducible representation of the group $G$ on the Hilbert space of each qudit, labelled by the index $k=1, \dots, L$. We assume that $d$ is such that there exists this irreducible representation. Of course, for a generic $d$, one could also consider a reducible representation, but this makes the discussion more cumbersome and the final leading asymptotic behaviour of the asymmetry is the same as the one we will find for the irreducible case.
We denote the generators of the group in this representation as $T_a^{(k)}$ with $a=1,\dots,\text{dim}\,G$ and $k=1,\dots, L$. For a semi-simple Lie algebra, $T_a^{(k)}$ are traceless $d\times d$ Hermitian matrices. It is possible to construct, for any fixed $k$, a basis of the semi-simple Lie algebra such that $\text{Tr}[T_a^{(k)} T_b^{(k)}] = \lambda \delta_{ab} $, where $\lambda$ is the Dynkin index of the chosen representation~\cite{DiFrancesco}. 

Our goal is to compute the average entanglement asymmetry associated with the global compact Lie group generated by the charges $Q_a = \sum_{k=1}^L T_a^{(k)}$ for $a =1,\cdots, \text{dim}\,G$. These charges are local, $Q_a=Q_{A, a}+Q_{B, a}$ and, therefore, the unitary representation $U_{g(\vec{\alpha})}=e^{i\vec{\alpha}\cdot \vec{Q}}$, where $\vec{\alpha}\in \mathbb{R}^{\dim G}$, of the elements of the group $g(\vec{\alpha})\in G$ in the full Hilbert space factorises as $U_{g(\vec{\alpha})}=U_{A, g(\vec{\alpha})}\otimes U_{B, g(\vec{\alpha})}$, as required.

Notice that the steps leading to the expression of the charge moments~\eqref{resEZ} for the $SU(2)$ do not actually rely on specific properties of the Lie group. The only difference is that now the Hilbert space of the total system has dimension $d^L$. This modifies the approximation of the Weingarten coefficients in~\eqref{Haar_avg} for large $L$ as $\text{Wg}(\sigma) \simeq d^{-nL} \delta_{\sigma,\text{Id}}$ and the interval over which the indices $s_j$ in the boundary states~\eqref{+state} and~\eqref{-state} run, which is now $s_j=1,\dots, d$. Therefore, Eq.~\eqref{resEZ} can be straightforwardly generalised to any compact Lie group as
\begin{equation}
\label{avgZnG}
\mathbb{E}[Z_n(\mathbf{g})] \simeq \frac{1}{d^{nL}} \sum_{\sigma \in S_n} \Big(\sum_{\{s_j = 1\}}^d \prod_{j=1}^{n} \delta_{s_j,s_{\sigma(j)}} \Big)^{L-\ell_A}\Big(\sum_{\{s_j = 1\}}^d \prod_{j=1}^{n} (\mathrm{U}_{g_j})_{s_{\sigma(j)},s_{j+1}}\Big)^{\ell_A},
\end{equation}
where $\mathrm{U}_{g}$ is the unitary irreducible representation of $g \in G$ on a local qudit Hilbert space given by $\mathrm{U}_g = \exp(i\alpha^a T_a)$, where here and in the following we use the convention that repeated algebra indices are summed over.

The structure of Eq.~\eqref{avgZnG}, apart from the different unitary representation $\mathrm{U}_g$, is the same as for the $U(1)$ and $SU(2)$ groups. Hence, the leading term at large $L$ is determined by the permutation maximising the number of undressed loops for $\ell_A<1/2$ --- the identity permutation~\eqref{+sigmaid}-\eqref{-sigmaid}--- and of dressed loops for $\ell_A>L/2$ --- the permutation $\nu(j)=j+1$~\eqref{+sigmacycl}-\eqref{-sigmacycl}. For a generic group $G$, the reason why the dominant term for $\ell_A>L/2$ is obtained by maximising the number of loops is less evident, but it will be made clear when computing the saddle-point approximation of the integral of the averaged charged moments. 
Temporarily leaving this clarification aside, we obtain that the leading term in Eq.~\eqref{avgZnG} is
\begin{equation}
\label{EZSUN}
\mathbb{E}[Z_n(\mathbf{g})] \stackrel{L \to \infty}{\simeq} \begin{cases}
d^{(1-n)\ell_A}, & \ell_A < L/2 \\
d^{(1-n)L} d^{-\ell_A} \big(\prod_{j=1}^{n-1}\text{Tr}[\mathrm{U}_j]^{\ell_A} \big) \text{Tr}[\mathrm{U}^{\dagger}_{n-1} \cdots \mathrm{U}^{\dagger}_{1}]^{\ell_A}, & \ell_A > L/2
\end{cases}.
\end{equation}
To obtain the average Rényi asymmetry, we need to plug this result into Eq.~\eqref{def_asym_chargedmom} and perform the integral. For values $\ell_A<L/2$, we have $\mathbb{E}[Z_n(\mathbf{g})] = \mathbb{E}[Z_n(\mathbf{e})]$, which immediately implies that $\mathbb{E}[\Delta S^{(n)}_A] = 0$ in agreement with the Hayden-Preskill decoupling inequality~\eqref{dec_ineq}.

For $\ell_A>L/2$, we need to analyse the behaviour of the integral~\eqref{def_asym_chargedmom} as $\ell_A \to \infty$ with fixed ratio $\ell_A/L$. We can apply a saddle point approximation as we did in the $SU(2)$ case. To this end, we need to find the subgroup of elements of $G$ that commute with all the group; this is its centre,  $Z(G) = \{h \in G \,|\, hgh^{-1} = g\,\, \forall g \in G\}$. For these points, as we are going to prove now, the charged moments~\eqref{charged_mom} are $\mathbb{E}[Z_n(\mathbf{h})]=\mathbb{E}[Z_n(\mathbf{e})]$, where $\mathbf{h} = (h_1,\cdots, h_{n-1}) \in Z(G)^{n-1}$. Since we are assuming that $\mathrm{U}_g$ is an irreducible representation, Schur's lemma ensures that for any element $h$ of the centre, $\mathrm{U}_h$ is proportional to the identity, i.e. $\mathrm{U}_h = \eta_h \mathbb{I}$. If we further assume that $G$ is a semi-simple compact Lie group, then the centre $Z(G)$ is necessarily a finite Abelian subgroup, see e.g.~\cite{h-72} for more details. This implies that $\mathrm{U}_h^p = \mathbb{I}$ for some integer $p$ and hence $\eta_h$ is a root of the unity, $\eta_h = e^{i \varphi(h)}$. 
Since $\mathrm{U}_h = e^{i \varphi(h)} \mathbb{I}$, it is clear from equation~\eqref{EZSUN} that, for any choice of $\mathbf{h}$, $ \mathbb{E}[Z_n(\mathbf{g} \cdot \mathbf{h})] = \mathbb{E}[Z_n(\mathbf{g})]$ where $\mathbf{g} \cdot \mathbf{h} = (g_1\cdot h_1,\cdots,g_{n-1}\cdot h_{n-1})$. As a result, the integrand in Eq.~\eqref{def_asym_chargedmom} has a saddle point at each element $\mathbf{h}\in Z(G)^{n-1}$ and the behaviour of the integrand around a neighbourhood of all these saddle points is the same.
Therefore, we can approximate the integral taking a neighbourhood $I(\vec{0})$ around the saddle point at the identity $\mathbf{g}=\mathbf{e}$, which corresponds in local coordinates to $\vec{\alpha}_j=\vec{0}$, $j=1, \dots, n-1$,  and multiply by the number of saddle points $|Z(G)|^{n-1}$,  obtaining
\begin{equation}
\label{saddleSUN}
\frac{\mathbb{E}[\text{Tr}[\rho_{A,G}^n]]}{\mathbb{E}[\text{Tr}[\rho_A^n]]} \simeq \frac{|Z(G)|^{n-1}}{(\text{vol}\,G)^{n-1}} \int_{I(\vec{0})} \text{d}\mathbf{g} \exp \left[\ell_A\left( \sum_{j=1}^{n-1} \log \text{Tr}[\mathrm{U}_j] + \log \text{Tr}[\mathrm{U}^{\dagger}_{n-1} \cdots \mathrm{U}^{\dagger}_{1}]\right)\right].
\end{equation}
In this expression, ${\rm d}\mathbf{g}$ stands for the Haar measure of the group $G$ in the local coordinates, ${\rm d}\mathbf{g}=\mu(\vec{\alpha}_1)\cdots\mu(\vec{\alpha}_{n-1}){\rm d}\vec{\alpha}_1\cdots {\rm d}\vec{\alpha}_{n-1}$.
Using the properties $\text{Tr}[T_a^{(k)}] = 0$ and $\text{Tr}[T_a^{(k)} T_b^{(k)}] = \lambda \delta_{ab}$, one can straightforwardly prove the following quadratic expansions around $\vec{\alpha}_j=\vec0$
\begin{equation}
\label{expTrU1}
\log \text{Tr}[\mathrm{U}_j] \simeq \log d - \frac{\lambda}{2 d} \alpha^a_j \alpha^a_j,   
\end{equation}
\begin{equation}
\label{expTrU2}
\log \text{Tr}[\mathrm{U}^{\dagger}_{n-1} \cdots \mathrm{U}^{\dagger}_{1} ] \simeq \log d - \frac{\lambda}{2d} \sum_{i,j = 1}^{n-1} \alpha^a_i \alpha^a_j.
\end{equation}
We can now clarify why the dominant contribution to the averaged charged moments is given by the permutation which maximises the number of dressed loops. The constant term in the expansions~\eqref{expTrU1} and~\eqref{expTrU2} implies that each trace $\text{Tr}[\mathrm{U}_j]$, which is precisely the contribution of a dressed loop, contributes with a factor $d^{\ell_A}$ to the integral while the quadratic terms in $\alpha_j^a$, once they are integrated, go to zero algebraically in $\ell_A$. 
As a result, the dominant contribution corresponds to the permutation that gives the maximum number of factors $\text{Tr}[\mathrm{U}_j]$ in the charged moments, i.e. maximises the number of dressed loops. 

Plugging the quadratic expansions~\eqref{expTrU1} and~\eqref{expTrU2} into Eq.~\eqref{saddleSUN} and doing a saddle point approximation, the $[\dim G(n-1)]$-dimensional integral decouples into $\dim G$ identical $(n-1)$-dimensional Gaussian integrals, similar to the one appearing in Eq.~\eqref{eq:gauss_int_su2} for the $SU(2)$ group,
\begin{equation}
\frac{\mathbb{E}[\text{Tr}[\rho_{A,G}^n]]}{\mathbb{E}[\text{Tr}[\rho_A^n]]} \simeq \left[\frac{|Z(G)|\, \mu(0)}{\text{vol}\,G}\right]^{n-1} \left[\int_{\mathbb{R}^{n-1}} \text{d}\alpha_1\cdots {\rm d}\alpha_{n-1} \,\, e^{-\frac{\ell_A \lambda}{2 d}(\sum_{j=1}^{n-1} \alpha_j^2 + \sum_{j,k=1}^{n-1} \alpha_j\alpha_k )}\right]^{\dim\,G},
\end{equation}
where $\mu(0) \equiv \lim_{\Vec{\alpha} \to \vec{0}} \mu(\vec{\alpha})$.
Finally, by applying the standard formulas of Gaussian integrals, we obtain the following result for the average entanglement asymmetry for a semi-simple compact Lie group $G$ in large systems,
\begin{equation}
\label{typvNEA_G}
    \mathbb{E}[\Delta S_A^{(n)}] \simeq \begin{cases}
        0, & \ell_A<L/2, \\
 \frac{\text{dim}\,G}{2} \log \Bigg[\frac{\ell_A \lambda}{2d \pi} \Bigg(\frac{\text{vol}\, G}{|Z(G)| \, \mu(0)}\Bigg)^{\frac{2}{\text{dim}\,G}} n^{\frac{1}{n-1}}\Bigg], & \ell_A>L/2.
    \end{cases}
\end{equation}
Taking the limit $n\to1$, we find that the von Neumann entanglement asymmetry is
\begin{equation}
\label{avgDSgeneralG}
    \mathbb{E}[\Delta S_A] \simeq  \frac{\text{dim}G}{2}\log \left[\frac{\ell_A \lambda}{2d \pi} \left(\frac{\text{vol}\, G}{|Z(G)| \, \mu(0)}\right)^{\frac{2}{\text{dim}\,G}}\right] +  \frac{\text{dim}G}{2}
       , \,\,\,\,  \ell_A > L/2.
\end{equation}

This is the main result of this work. 
As established in Ref.~\cite{ampc-24} for the $U(1)$ group and in the previous section for $SU(2)$, the entanglement asymmetry of any compact, semi-simple Lie group jumps from zero to a finite value at $\ell_A=L/2$ for large system sizes. 
One can check, using the same arguments that we employed in Sec.~\ref{sec_abelian} for the $U(1)$ group, that the variance of the entanglement asymmetry vanishes in the limit $L\to\infty$. Therefore, the result in Eq.~\eqref{avgDSgeneralG} is also the typical entanglement asymmetry. In the regime $\ell_A>L/2$, it diverges logarithmically in the subsystem size with a prefactor which only depends on the dimension of the group considered. 
The same scaling for the entanglement asymmetry of a compact, semi-simple Lie group has also been established in substantially different contexts. Such scaling was found in Refs.~\cite{chmp-20,ch-22} in the vacuum of quantum field theories. Other examples include
the ground state of spin chains (either critical or not) where the symmetry is completely broken and no continuous subgroups remain as symmetries of the state~\cite{cv-23,fadc-24, lmac-24}. Note that in those ground states, the entanglement entropy satisfies either an area-law or grows logarithmically with the subsystem~\cite{afov-08}. In our case, instead, the Haar random states exhibit, on average, an extensive entanglement entropy, according to the Page curve~\eqref{eq:page_curve}.
Therefore, the logarithmic scaling $(\text{dim}\,G/2) \log \ell_A$ is a very universal result that only depends on the way the state breaks the symmetry but not on its entanglement content.

\subsection{The explicit example of the $SU(N)$ group.}
As an example, we specialise the result in Eq.~\eqref{typvNEA_G} to the semi-simple compact Lie Group $SU(N)$. We consider a system of qudits of dimension $d=N$ and, consequently, $SU(N)$ acts locally as the fundamental representation. In this case, we have that $\text{dim}\, SU(N) = N^2-1$ and $|Z(SU(N))| = N$. We also need its volume $\text{vol}\,SU(N)$ and the value of its Haar measure at $\vec{\alpha}=\vec{0}$, $\mu(0)$. To determine $\mu(0)$, we recall that the Haar measure of $SU(N)$ is induced by the Killing form on the Lie algebra $\mathfrak{su}(N)$, $K(X, Y)$ with $X, Y\in \mathfrak{su}(N)$, as
\begin{equation}
\label{Kinducedmeasure}
    \mu(\vec{\alpha}) = \sqrt{\text{det}\,\mathcal{K}_{ab}(\vec{\alpha})} ,
\end{equation}
where
\begin{equation}
\label{Killingform}
    \mathcal{K}_{ab}(\vec{\alpha}) \equiv K(i g^{-1}(\vec{\alpha}) \partial_a g(\vec{\alpha}),i g^{-1}(\vec{\alpha}) \partial_bg(\vec{\alpha})),
\end{equation}
and $\partial_a \equiv \partial_{\alpha^a}$.
For the Killing form, we use the normalisation convention of Ref.~\cite{DiFrancesco}, i.e. $K(X,Y) = \frac{1}{2g}\text{Tr}[\text{ad}X\text{ad}Y]$, where $g$ is the dual Coxeter number and $\text{ad} X$ the adjoint representation of $X$. The dual Coxeter number of $SU(N)$ is $g = N$, so within our convention, the Killing form reads $K(X,Y) = \text{Tr}[X^{\dagger}Y]$. We have now all the necessary ingredients to compute $\mu(0)$. Using the $SU(N)$ Killing form in~\eqref{Killingform}, we obtain $\mathcal{K}_{ab}(\vec{\alpha}) = \text{Tr}[(\partial_a g(\vec{\alpha}))^{\dagger} \partial_b g(\vec{\alpha})]$. Expanding in series $g(\vec{\alpha}) = \exp(i\alpha^a T_a)$ around $\vec{\alpha}=\vec{0}$, we find
\begin{equation}
    \mathcal{K}_{ab}(\vec{\alpha}) = \text{Tr}[\partial_a( \mathbb{I}-i\alpha^c T_c + \mathcal{O}(\alpha^2) )\partial_b(\mathbb{I}+i\alpha^d T_d + \mathcal{O}(\alpha^2))].
\end{equation}
If we now calculate the derivatives in the expression above, we get $\mathcal{K}_{ab}(\vec{\alpha}) \simeq \text{Tr}[T_a T_b] + \mathcal{O}(\vec\alpha)$. Therefore, taking into account that $\text{Tr}[T_aT_b]=\lambda \delta_{ab}$, in the limit $\vec{\alpha}\to 0$, we obtain $\mathcal{K}_{ab}(\vec{0}) = \lambda \delta_{ab}$. Finally, using this result in Eq.~\eqref{Kinducedmeasure}, we conclude that the $SU(N)$ Haar measure at $\vec{\alpha}=\vec{0}$ is
\begin{equation}
    \mu(0) = \lambda^{\text{dim}\,SU(N)/2}.
\end{equation}
The volume of the group, $\text{vol}\,SU(N)$, is largely discussed in the literature. In our normalisation convention for the Killing form, it is given by~\cite{cdm-21}
\begin{equation}
\label{volSUN}
    \text{vol}\, SU(N) = \sqrt{N} \frac{(2\pi)^{\frac{N^2+N}{2}-1}}{G(N+1)},
\end{equation}
where $G(N+1) = \prod_{j=1}^{N-1}j!$ is the Barnes $G$ function.

Inserting all the previous ingredients in Eq.~\eqref{avgDSgeneralG}, the average R\'enyi entanglement asymmetry for the $SU(N)$ group in the regime $\ell_A>L/2$ is thus given by
\begin{equation}
\label{finitevalueEDS_SUN}
    \mathbb{E}[\Delta S^{(n)}_A] \simeq \frac{N^2-1}{2} \log \left[\frac{\ell_A}{2 N \pi} \left( \frac{\text{vol} \,SU(N)}{N}\right)^{\frac{2}{N^2-1}} n^{\frac{1}{n-1}}\right], \quad \ell_A>L/2.
\end{equation}

In the following, we provide some numerical checks of this result in the special case of the group $SU(3)$. First, in Fig.~\ref{fig:checkZ3SU3}, we verify that Eq.~\eqref{EZSUN} is the correct leading asymptotic behaviour of the average charged moments. To do so, we can calculate numerically the $n=3$ average charged moments $\mathbb{E}[Z_3(\mathbf{g})]$ for $SU(3)$ using Eq.~\eqref{avgZnG} and considering all possible permutations in $S_3$. We can then compare the result (solid lines) with the ones obtained using the approximation in Eq.~\eqref{EZSUN}, where only the leading permutation is taken into account (dashed lines). In both panels, we take a fixed total system size ($L=6$ on the left and $L=16$ on the right) and we plot the charged moments as a function of a parameter $\kappa$ of the local coordinates for increasing subsystem sizes $\ell_A$. As $\ell_A$ increases, the agreement between the large system approximation~\eqref{EZSUN} and~\eqref{avgZnG} improves.
Moreover, in the left panel, we further verify the validity of Eq.~\eqref{avgZnG} by checking it against the average $Z_3(\mathbf{g})$ over a finite ensemble of Haar random states computed using exact diagonalisation (symbols).

\begin{figure}[t]
\begin{subfigure}
    \centering
    \includegraphics[width=.5\linewidth]{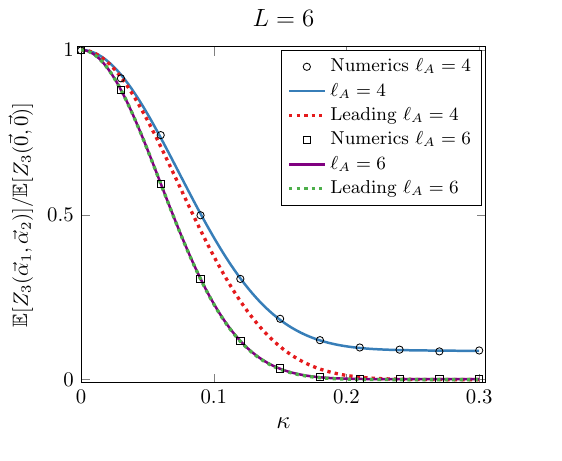}
\end{subfigure}
\begin{subfigure}
    \centering
    \includegraphics[width=.5\linewidth]{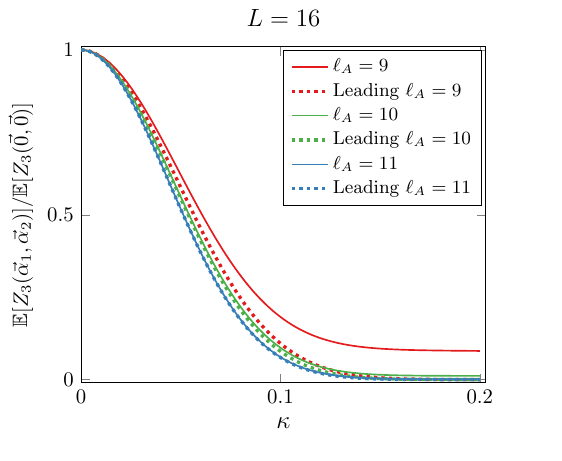}
\end{subfigure}
\caption{Average $n=3$ charged moment $\mathbb{E}[Z_3(\vec{\alpha}_1,\vec{\alpha}_2)]$ for the $SU(3)$ group. 
We take a system of $L=6$ (left panel) and $L=16$ (right panel) qutrits and different subsystem sizes $\ell_A$. We plot the charged moments as a function of the local coordinate parameter $\kappa$, chosen as $\vec{\alpha}_1 = \kappa \cdot (5.51,0.17,0.48,3.55,4.62,2.25,6.16,5.82)$ (randomly generated) and $\vec{\alpha}_2 = (0,0,\kappa,0,0,0,0,0)$. 
In both panels, the solid curves are the analytical prediction~\eqref{avgZnG} at finite $\ell_A$, while the dashed ones correspond to the large-system asymptotics~\eqref{EZSUN}. In the left panel, the symbols are the value obtained calculating the charged moments with exact diagonalisation and averaging over $10^2$ Haar random states. The exponential growth of the Hilbert space as $3^L$ makes it difficult to reach larger system sizes using exact diagonalisation and, for this reason, the right panel does not show numerics.}
\label{fig:checkZ3SU3}
\end{figure}

In Fig.~\ref{fig:asym2SU3check}, we check the saddle point prediction~\eqref{finitevalueEDS_SUN} in the case of the  $n=2$ Rényi asymmetry for $SU(3)$. It is possible to compute $\mathbb{E}[\Delta S^{(2)}_A]$ by numerically integrating over the $SU(3)$ group the averaged charged moments $\mathbb{E}[Z_2(\mathbf{g})]$ using the formula~\eqref{EZSUN} for them. To perform the numerical integration over the $SU(3)$ group, we employ the Euler angle parameterisation of the $d=3$ fundamental representation of $SU(3)$, as discussed in e.g.~\cite{byrd-97},
\begin{equation}
    \mathrm{U} = e^{i \lambda_3 \alpha} e^{i \lambda_2 \beta} e^{i \lambda_3 \gamma} e^{i \lambda_5 \theta} e^{i \lambda_3 a} e^{i \lambda_2 b} e^{i \lambda_3 c} e^{i \lambda_8 \phi},
\end{equation}
where $\lambda_i$ are the Gell-Mann matrices normalised as $\text{Tr}[\lambda_a \lambda_b] = 2 \delta_{ab}$ and 
\begin{equation}
0\leq \alpha, \gamma, a, c<\pi,\quad 0\leq \beta, b, \theta\leq \pi/2,\quad 0\leq \phi <2\pi.
\end{equation}
In this parameterisation, the Haar measure is 
\begin{equation}
    {\rm d}\mu = \sin (2\beta) \sin(2b) \sin (2\theta) \sin^2(\theta) \, {\rm d}\alpha \, {\rm d}\beta \, {\rm d}\gamma \, {\rm d}\theta \, {\rm d}a \, {\rm d}b \, {\rm d}c \, {\rm d}\phi.
\end{equation}
The $8$-dimensional integral of $\mathbb{E}[Z_2(\mathbf{g})]$ has been computed numerically using Monte Carlo method. In Fig.~\ref{fig:asym2SU3check}, we compare the results obtained numerically with the large $\ell_A$ saddle-point prediction \eqref{finitevalueEDS_SUN} for the case in which $\ell_A=L$.

\begin{figure}[t]
\includegraphics[width=0.7\textwidth]{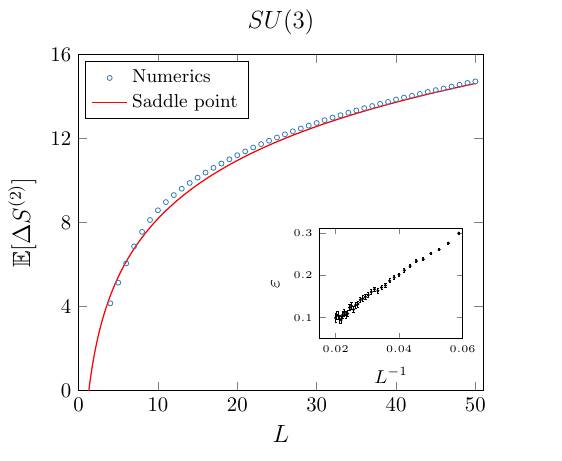}
\centering
\caption{Average $SU(3)$ $n=2$ Rényi entanglement asymmetry $\mathbb{E}[\Delta S^{(2)}]$ taking $\ell_A = L$ and varying the total system size $L$. The solid line is the asymptotic analytical prediction obtained by a saddle point approximation in Eq.~\eqref{finitevalueEDS_SUN}. 
The symbols were obtained by numerically integrating as in Eq.~\eqref{def_asym_chargedmom} the average charged moments in Eq.~\eqref{avgZnG}. The integration was performed over the $SU(3)$ group, using Monte Carlo method and employing the Euler angle parameterisation described in detail in the main text.
The statistical uncertainty for all the points is smaller than the size of the marker and the relative error is approximately $\lesssim10^{-3}$. Inset:~Difference $\varepsilon$ between the numerical integration and the saddle point approximation~\eqref{finitevalueEDS_SUN} as a function of $L^{-1}$.}
\label{fig:asym2SU3check} 
\end{figure}

\end{section}
\begin{section}{Conclusions}
\label{sec_concl}

In this work, we studied the breaking of a generic global internal symmetry in Haar random states. To this end, we employed the entanglement asymmetry, an entanglement-based observable that quantifies how much a symmetry is broken in a part of an extended quantum system. 
Our findings generalise the recent results presented in Ref.~\cite{ampc-24} for the $U(1)$ group, extending them to encompass any compact, semi-simple Lie group.
In the large system-size $L$ limit, the average entanglement asymmetry is zero when the subsystem size $\ell_A$ satisfies $\ell_A<L/2$. 
At $\ell_A=L/2$, it suddenly jumps to a finite value, meaning that there is a sharp transition in the subsystem $A$ from a state that respects any symmetry to a non-symmetric state. 
The vanishing of the entanglement asymmetry for any group when $\ell_A < L/2$ can be understood through the Hayden-Preskill decoupling inequality~\eqref{dec_ineq}. This inequality states that, for $\ell_A < L/2$, the reduced density matrix of subsystem  A  is, on average, exponentially close to the (normalised) identity matrix, which is inherently symmetric under any group. However, the decoupling inequality does not explain the behaviour of the entanglement asymmetry when $\ell_A\geq L/2$, particularly the abrupt discontinuity observed at $\ell_A=L/2$. 
For $\ell_A>L/2$, the entanglement asymmetry grows logarithmically in $\ell_A$ with a coefficient fixed by the dimension of the group. In this regime, we also  calculated the exact value of the $\ell_A$-independent term for any value of the R\'enyi index, which is expressed in terms of a few specific parameters of the group. 
As an illustration, we explicitly computed it for the $SU(N)$ group. We also discussed the size of the fluctuations of the asymmetry around the average value. We showed that they go to zero as the system size $L$ increases and, therefore, the average entanglement asymmetry that we computed is also the typical one.

Our results have relevance in multiple contexts, as the states we are considering can be thought as representatives of excited eigenstates of chaotic many-body Hamiltonians. As we already mentioned, our system of random qudits is the same originally studied by Page in~\cite{p-93, p-93-1} to model the evaporation of a black hole. In that case, we can think that the subsystem $A$ is the radiation emitted while $B$ is the black hole. Before it starts to evaporate, all the qudits are in $B$ and, therefore, $\ell_A=0$. As the black hole evaporates, the qudits are transferred to $A$, mimicking the emission of radiation, until $\ell_A=L$ when the black hole is fully evaporated. In this process, we can identify time as $t\propto \ell_A$. As already discussed in Ref.~\cite{ampc-24}, our results imply that, if the black hole is initially in a state that breaks any symmetry and information is preserved during its evaporation, then the radiation emitted is in a fully symmetric state until the halfway point of the process, $\ell_A=L/2$, i.e. the Page time. 
At that moment, not only does the entanglement entropy of the radiation begin to decrease, following the Page curve~\eqref{eq:page_curve}, but the state also abruptly loses any symmetry, mirroring the symmetry breaking that initially occurred in the black hole.
Thus, until the Page time, studying the radiation alone does not reveal whether the black hole breaks any symmetry.
We remark that, in our setup, the black hole’s state prior to evaporation breaks any global symmetry by assumption. The mechanism behind the breaking of global symmetries due to quantum gravity effects
is outside the scope of this model.
Our conclusions are derived under fairly general assumptions and, much like the arguments of Wigner and Dyson for nuclei, are likely to hold in specific models of quantum gravity based on principles of typicality.
For example, the sudden transition in the radiation from a symmetric to a non-symmetric state has been also observed in a particular theory of quantum gravity within the replica wormhole formalism~\cite{cl-21}. In that case, the jump discontinuity in the entanglement asymmetry can be traced back to the appearance at the Page time of an ``island''~\cite{aemm-19,g-24}, that is a region inside the black hole that contributes to the radiation's entanglement entropy. See also Refs.~\cite{hs-21, hiz-21} for related discussions on the explicit breaking of symmetries during black hole evaporation.

Another relevant area is the thermalisation of generic isolated quantum systems. The Haar random states studied here describe their long-time behaviour in the absence of any conserved quantity. According to our findings, in the thermodynamic limit, subsystems of size $\ell_A<L/2$ relax to a fully symmetric state under any group, while the larger ones, $\ell_A>L/2$ converge to a state that does not respect any symmetry. This is important, for example, in the analysis of anomalous relaxation phenomena, such as the quantum Mpemba effect~\cite{amc-23}, already explored for the $U(1)$ group in random circuits~\cite{tcd-24,lzyz-24}, which could be extended to larger, non-Abelian symmetries.

In the Haar random state ensemble, the behaviour of the average entanglement asymmetry for $\ell_A< L /2$ is highly constrained by the Hayden-Preskill decoupling inequality. A natural prospect is to consider other ensembles of random states endowed with some additional structures such as a conserved symmetry group. In this case, the decoupling inequality might be spoiled, giving rise to a phenomenology different from the scenario put forward in the present work.

\textbf{Acknowledgments.} We thank Sara Murciano and Lorenzo Piroli for collaboration in related topics. We thank Marina Huerta for bringing references~\cite{chmp-20, chmp-21, magan-21, bcklm-24} to our attention. PC and FA acknowledge support from ERC under Consolidator Grant number 771536 (NEMO) and from European Union - NextGenerationEU, in the framework of the PRIN Project HIGHEST number 2022SJCKAH$\_$002. 

\end{section}

\end{document}